\documentclass[prb,twocolumn,aps,superscriptaddress,showpacs]{revtex4-2}

\usepackage{amsmath,amssymb}
\usepackage{graphicx}
\usepackage{xcolor}
\usepackage{graphicx}
\DeclareGraphicsExtensions{.pdf,.eps,.png,.jpg} \graphicspath{{./figs/}}
\usepackage{dcolumn}
\usepackage{bm}
\usepackage{tabularx}
\usepackage{upgreek}
\usepackage{multirow}
\usepackage{epstopdf}
\usepackage{gensymb}

\usepackage{amsmath}%
\usepackage{amsfonts}%
\usepackage{amssymb}%
\usepackage[latin1]{inputenc}
\usepackage{graphicx}
\usepackage{epstopdf}
\usepackage{color}
\usepackage{ulem}
\usepackage[english]{babel}
\usepackage{natbib}
\usepackage[colorlinks=true, citecolor=blue, linkcolor=blue, urlcolor=blue]{hyperref}
\usepackage{placeins}
\usepackage{booktabs}
\usepackage{enumitem}
\usepackage{subfigure}
\usepackage{multirow}
\usepackage{inputenc}
\usepackage{braket}
\usepackage{svg}
\usepackage{dcolumn}
\usepackage{bm}
\usepackage{siunitx}
\usepackage{mathtools}

\begin{document}

\title{Dilution of the magnetic lattice in the Kitaev candidate $\alpha$-RuCl$_3$  by Rh$^{3+}$ doping}

\author{G. Bastien}
\thanks{These two authors contributed equally to the work.}
\email{gael.bastien@mag.mff.cuni.cz}
\affiliation{Leibniz IFW Dresden, Institute of Solid State Research, 01069
Dresden, Germany}
\affiliation{Charles University, Faculty of Mathematics and Physics, Department of Condensed Matter Physics, Prague, Czech Republic}
\author{E. Vinokurova}
\thanks{These two authors contributed equally to the work.}
\affiliation{Leibniz IFW Dresden, Institute of Solid State Research, 01069
Dresden, Germany}
\affiliation{Institut f\"ur Festk\"orper- und Materialphysik, Technische
Universit\"at Dresden, 01062 Dresden, Germany}
\author{M. Lange}
\affiliation{Leibniz IFW Dresden, Institute of Solid State Research, 01069
Dresden, Germany}
\author{K. K. Bestha}
\affiliation{Leibniz IFW Dresden, Institute of Solid State Research, 01069
Dresden, Germany}
\author{L.~T.~Corredor}
\affiliation{Leibniz IFW Dresden, Institute of Solid State Research, 01069
Dresden, Germany}
\author{G.~Kreutzer}
\affiliation{Leibniz IFW Dresden, Institute of Solid State Research, 01069
Dresden, Germany}
\author{A.~Lubk}
\affiliation{Leibniz IFW Dresden, Institute of Solid State Research, 01069
Dresden, Germany}
\affiliation{Institut f\"ur Festk\"orper- und Materialphysik and W\"urzburg-Dresden Cluster of Excellence ct.qmat, Technische
Universit\"at Dresden, 01062 Dresden, Germany}
\author{Th.~Doert}
\affiliation{Faculty of Chemistry and Food Chemistry, Technische Universit\"at Dresden, 01062 Dresden, Germany}
\author{B. B\"uchner}
\affiliation{Leibniz IFW Dresden, Institute of Solid State Research, 01069
Dresden, Germany}
\affiliation{Institut f\"ur Festk\"orper- und Materialphysik and W\"urzburg-Dresden Cluster of Excellence ct.qmat, Technische
Universit\"at Dresden, 01062 Dresden, Germany}
\author{A. Isaeva}
\email{a.isaeva@uva.nl}
\affiliation{Leibniz IFW Dresden, Institute of Solid State Research, 01069
Dresden, Germany}
\affiliation{Van der Waals-Zeeman Institute, Department of Physics and Astronomy, University of Amsterdam, Science Park 094, 1098 XH Amsterdam, The Netherlands}
\author{A. U. B. Wolter}
\affiliation{Leibniz IFW Dresden, Institute of Solid State Research, 01069
Dresden, Germany}

\date{\today}

\begin{abstract}

Magnetic dilution of a well-established Kitaev candidate system is realized in the substitutional Ru$_{1-x}$Rh$_x$Cl$_3$ series ($x = 0.02-0.6$). Optimized syntheses protocols yield uniformly-doped single crystals and polycrystalline powders that are isostructural to the parental $\alpha$-RuCl$_3$ as per X-ray diffraction. The Rh content $x$ is accurately determined by the quantitative energy-dispersive X-ray spectroscopy technique with standards. We determine the magnetic phase diagram of Ru$_{1-x}$Rh$_x$Cl$_3$ for in-plane magnetic fields from magnetization and specific-heat measurements as a function of $x$ and stacking periodicity, and identify the suppression of the magnetic order at $x \approx 0.2$ towards a disordered phase, which does not show any clear signature of freezing into a spin glass. Comparing with previous studies on the substitution series Ru$_{1-x}$Ir$_x$Cl$_3$, we propose that chemical pressure would contribute to the suppression of magnetic order especially in Ru$_{1-x}$Ir$_x$Cl$_3$ and that the zigzag magnetic ground state appears to be relatively robust with respect to the dilution of the Kitaev--$\Gamma$--Heisenberg magnetic lattice. We also discovered a slight dependence of the magnetic properties on thermal cycling, which would be due to an incomplete structural transition.

\end{abstract}


\maketitle

\section{Introduction}

Frustrated magnetism attracts much attention as a cradle of competing magnetic interactions and, possibly, quantum spin liquid phases~\cite{Kitaev2006,Balents2010,Winter2017,Takagi2019}. One way of tweaking the properties of frustrated magnets is to induce a structural disorder in the sublattice of the magnetic ions such as vacancies~\cite{Maryasin2013,Dey2020,Kimchi2018}. This can be achieved via the partial substitution of the magnetic atoms by their non-magnetic counterparts. Such dilution of magnetic moments reduces the connectivity of the network of magnetic ions with substantial disorder. The study of the effect of structural disorder in frustrated magnets is motivated by the recent discovery of many quantum spin liquid candidate materials with a large amount of structural disorder such as the Kitaev magnets H$_3$LiIr$_2$O$_6$, Cu$_2$IrO$_3$ and OsCl$_3$ \cite{Kitagawa2018,Yadav2018,Kao2021,Kenney2019,Kataoka2020}.

Here, we consider the case of a Kitaev-Heisenberg-$\Gamma$ magnet on a honeycomb lattice, also called extended Kitaev model, based on the $j_{eff}=1/2$ Mott insulator $\alpha$-RuCl$_3$. In this system, the strong spin-orbit coupling combined with edge-sharing octahedra geometry induces a bond directional Ising-like Kitaev interaction ($K$) and an off-diagonal magnetic interaction $\Gamma$, besides the isotropic magnetic Heisenberg interaction ($J$)~\cite{Kitaev2006,Jackeli2009,Banerjee2016,Winter2017,Janssen2017}. A remarkable property of $\alpha$-RuCl$_3$ is its proximity to the pure Kitaev model ($J=0$, $\Gamma=0$) with $K/J\approx 10$ and $-K/\Gamma \approx 2$~\cite{Winter2017,Janssen2017}. The Kitaev model has attracted very strong attention in the past few years, since it is exactly solvable and it harbors a quantum spin liquid ground state with Majorana fermions as magnetic excitations~\cite{Kitaev2006}. Despite its proximity to the Kitaev model, $\alpha$-RuCl$_3$ accommodates an antiferromagnetic ground state with antiferromagnetic zig-zag order in the honeycomb plane~\cite{Johnson2015, Sears2015}. However, numerous measurements such as specific heat, NMR and microwave absorption have shown the proximity to the Kitaev spin-liquid state and the possible occurrence of Majorana Fermions as magnetic excitation in the paramagnetic state~\cite{Do2017, Jansa2018, Wellm2018}.

$\alpha$-RuCl$_3$ 
crystallizes in a monoclinic unit cell (sp.~gr.~$C2/m$, No.~12) and has a layered structure with an almost closest cubic chlorine packing ($ABC$ stacking periodicity). The Ru atoms are octahedrally coordinated by the Cl atoms and form honeycomb-like arrangements in the $ab$ plane. Adjacent RuCl$_3$ layers are separated by van der Waals gaps which renders the material prone to stacking faults.
Crystals with minimal amount of stacking faults show an antiferromagnetic ordering at $T_{\mathrm N}\backsimeq$~7\,K compatible with the monoclinic lattice symmetry~\cite{Johnson2015, Cao2016}. Crystals with a larger amount of stacking faults, i.e., an increasing portion of the {\it AB} stacking periodicity (see~Fig.~1 in~\cite{Cao2016} for a structure diagram), order at $T_{\mathrm N}\backsimeq$14~K~\cite{Cao2016}. In both cases, there is an in-plane zigzag spin structure in each individual honeycomb layer. There is also an elusive experimental indication that $\alpha$-RuCl$_3$ transforms into a rhombohedral modification at low temperature (sp.~gr.~$R\bar{3}$, No.~148). While this structure transition was evidenced at $T_s$=160\,K upon warming,~\cite{Kubota2015}, the transition upon cooling is strongly crystal-dependent and was reported to occur either at 60\,K~\cite{Baek2017, He2018} or at 140\,K~\cite{Kubota2015, Widmann2019, Gass2020}, or even as a two-step process~\cite{Park2016, He2018}. A similar structural phase transition was reported earlier for CrCl$_3$ between 298~K and 225~K~\cite{Morosin1964}, but is still debated in the case of $\alpha$-RuCl$_3$~\cite{Park2016,He2018,Widmann2019,Gass2020,Johnson2015}. In contrast to the monoclinic structure in $C2/m$, the rhombohedral structure has the hexagonal closest chlorine packing ($AB$ stacking periodicity) and is expanded in the $c$ direction ($c_{rhomb} = 3c_{mon} \sin{\beta}$), so that it contains three trilayers per unit cell. For a very informative visual representation of the relation between the different space groups proposed for $\alpha$-RuCl$_3$ we refer the reader to~Fig.~1 in~\cite{Mu2022}. It is enticing to assume that crystals with $T_{\mathrm N}\backsimeq$14~K also undergo a structural transition at $T_s$ and have a regular $AB$ stacking sequence, but observations of these two critical temperatures do not always correlate across the current literature.
   
Various experimental studies have been carried out to test the stability of the magnetic order in $\alpha$-RuCl$_3$ and to search for ways to destabilize it towards the Kitaev quantum spin liquid. First attempts with an applied magnetic field in the basal $ab$ plane found a suppression of magnetic order through a quantum critical point at $\mu_0H_c\backsimeq 7$\,T developing towards a gapped magnetically polarized phase~\cite{Majumder2015,Kubota2015,Leahy2017,Baek2017,Wolter2017,Hentrich2019,Balz2019}. The possible occurrence of a quantum spin liquid in a narrow field interval around $\mu_0H_c$ was proposed from thermal Hall effect~\cite{Kasahara2018,Yokoi2021,Bruin2022} and neutron scattering~\cite{Balz2019,Zhao2022} and remains under debate. Hydrostatic pressure studies have shown a reduction of the Neel temperature~\cite{Wolf2022}, however, a first-order transition towards a valence bond crystal occurs at $0.1~$GPa before the suppression of the magnetically ordered state~\cite{Cui2017,Bastien2018,Biesner2018,Wolf2022}. Similarly, the chemical substitution of Ru$^{3+}$ by the magnetic cations Os$^{3+}$ also leads to the formation of spin-singlet dimers~\cite{Kataoka2022}. Finally, the chemical substitution of Ru by the magnetic ion Cr$^ {3+}$ indeed destabilizes the magnetic ground state around 0.05 Cr/f.u., however, instead of a quantum disordered phase the formation of a spin-glass state was observed ~\cite{Roslova2019,Bastien2019,HILLEBRECHT199770}. 
  
In this context, the dilution of the magnetic lattice by random substitution of the Ru$^ {3+}$ cations by non-magnetic counterparts appears as a remaining route to destabilize the magnetic ground state of $\alpha$-RuCl$_3$ towards a quantum disordered phase. The effect of vacancies was deeply studied theoretically in the pure Kitaev model~\cite{Willans2010,Das2016,Vojta2016,Sanyal2021,Kao2021,Nasu2021}. These studies have shown that the Kitaev spin liquid is robust against magnetic dilution  up to at least 0.2~vacancies per formula unit~\cite{Willans2010,Sanyal2021,Nasu2021} and that vacancies can be used to tune locally the magnetic density of states~\cite{Sanyal2021,Kao2021,Nasu2021}. For the case of a Kitaev-Heisenberg magnet with a long-range zigzag magnetic order as ground state, Monte-Carlo calculations indicated the realization of a spin-glass state with short-range zigzag magnetic correlation~\cite{Andrade2014}, in agreement with experimental results on the honeycomb iridates Li$_2$Ir$_{1-x}$Ti$_x$O$_3$ and Na$_2$Ir$_{1-x}$Ti$_x$O$_3$~\cite{Manni2014}. This is in contrast to the case of dilution of the magnetic lattice in Ru$_{1-x}$Ir$_x$Cl$_3$, where the suppression of the antiferromagnetic order towards a possible diluted quantum spin-liquid state was reported around $x=0.13$ without signatures for a spin-glass state~\cite{Lampen-Kelley2017,Do2018,Do2020,Baek2020}. 

Here, we introduce the substitutional series Ru$_{1-x}$Rh$_x$Cl$_3$ as another example for magnetic dilution in $\alpha$-RuCl$_3$. We optimized the crystal-growth protocol from~\cite{Do2018} to obtain both small and large crystals of Ru$_{1-x}$Rh$_x$Cl$_3$  ($x = 0.02-0.4$) as well as polycrystalline powders ($x = 0.1-0.6$). We report crystal-structure determination by X-ray diffraction and the chemical-composition by quantitative EDS (energy-dispersive X-ray spectroscopy) with standards. The magnetic characterization of the crystal by magnetization and specific heat measurements show that the magnetic ground state of $\alpha$-RuCl$_3$ is relatively robust against vacancy disorder and that it subsists until a substitution ratio around $x^*=0.2$ ($x^*$ is determined from calibrated EDX measurements, see section~III.A), where magnetic long-range order is suppressed. The comparison between the experimental results obtained in Ru$_{1-x}$Rh$_x$Cl$_3$ with previous results in Ru$_{1-x}$Ir$_x$Cl$_3$ allow us to distinguish between the intrinsic effect of the magnetic dilution and the additional effect of chemical pressure.

\section{Experimental details}

\subsection{Powder syntheses and crystal growth}

\begin{figure}[h]
\begin{center}
\includegraphics[width=1.0\linewidth]{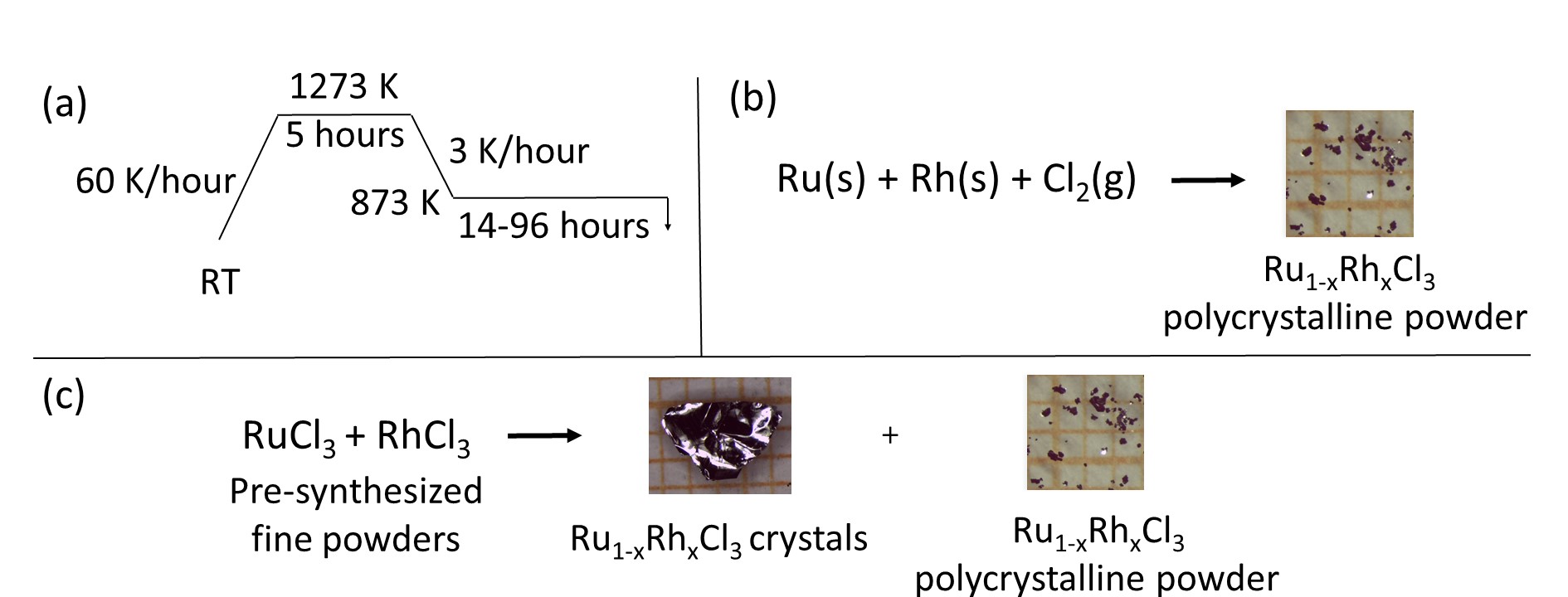}
\caption{(a) General temperature profile applied for the syntheses of Ru$_{1-x}$Rh$_x$Cl$_3$. (b) A schematic reaction of the elements and an optical image of the product. (c) A schematic reaction for the binaries and optical images of the products.}
    \label{tab:growth} 
\end{center}
\end{figure}

A series of Ru$_{1-x}$Rh$_x$Cl$_3$ specimens was obtained either from the elements or from the presynthesized $\alpha$-RuCl$_3$ \cite{binnewies2012} and RhCl$_3$ powders  \cite{barnighausen1964, Albrecht}. Further details of the syntheses optimizations are given in the Supporting Information, section~1. Fig.~\ref{tab:growth} sketches the protocols for obtaining polycrystalline powders of Ru$_{1-x}$Rh$_x$Cl$_3$ ($x=0.1-0.6$) and larger crystals of $1-5$ mm$^2$ size with  ($x=0.02-0.45$). The term ``polycrystalline powder'' refers to a collection of small platelet-like crystals of ca.\ 0.01--0.2 mm$^2$ in diameter (Fig.~\ref{tab:growth}b) that we used for powder X-ray diffraction studies and EDS. Large crystals were picked out for the magnetic studies. EDS experiments were also performed on pieces of these larger crystals to determine their chemical compositions.

\subsection{Scanning electron microscopy and energy-dispersive X-ray spectroscopy}

SEM images were taken on a Hitachi SU8020 microscope equipped with a X-Max$^{N}$ (Oxford) Silicon Drift Detector (SDD) with 2~kV  acceleration voltage and 5~$\mu$A  current. The chemical composition of Ru$_{1-x}$Rh$_x$Cl$_3$ was determined by calibrated energy-dispersive X-ray spectroscopy (EDS) on either the Hitachi instrument or on a high-resolution SEM
EVOMA 15 (Zeiss) equipped with a Peltier-cooled Si(Li) detector (Oxford Instruments) employing 30~kV acceleration voltage.
This voltage was set to generate the $K$ edge lines of Ru and Rh. Element
quantification was obtained from least-square fitting of edge models
(Ru-$K$, Rh-$K$, Cl-$K$) invoking k factor calibration from the
stoichiometric samples of similar chemical composition (RuCl$_3$ ($x=0$) and RhCl$_3$ ($x=1$)). The count rate from 5\,000\,000 to 50\,000\,000 counts and pulse processing time of 5~min were set in order to obtain a good peak-to-noise ratio. In addition, we tested quantification schemes involving the Ru-$L$ and Rh-$L$ edge lines, but their outcome did not appear reliable.

\subsection{X-ray diffraction}

Single-crystal X-ray diffraction (SCXRD) data were collected at 298~K on a STOE IPDS II with a STOE imaging plate detector using Mo-${K\alpha}$ radiation ($\lambda =0.71073$ \AA). The datasets were processed in the STOE X-Area software package. The crystal structure elucidation was performed in SHELXT~\cite{Sheldrick} and Olex2~\cite{Olex2} using the SHELXL~\cite{Sheldrick:fa3356} refinement package. For some crystals, low-temperature datasets were collected at 100~K on a four-cycle Supernova diffractometer from Rigaku-Oxford Diffraction with a hybrid photon counting detector using Mo-${K\alpha}$ radiation ($\lambda =0.71073$ \AA) and a graphite monochromator. Numerical absorption corrections were applied.

Powder X-ray diffraction (PXRD) data was collected using PANalytical X'Pert Pro or Empyrean diffractometers, both in the Bragg--Brentano geometry with a curved Ge(111) monochromator using Cu-${K\alpha_1}$ radiation ($\lambda$=1.54056 \AA) at room temperature. Si powder (National Institute of Standards and Technology, Gaithersburg, USA) with $a = 5.431179$~\AA~was mixed with powdered samples to be used as an internal standard for lattice parameters refinements. Polycrystalline powders with nominal $x=0, 0.1, 0.15, 0.3, 0.4, 1$ were measured in the angular range $5^{\circ}  \leq \theta \leq 90^{\circ}$ with a step size of $0.0066^{\circ}$ and time per step: 137.7~s. The sample  $x=0.5$ was measured with a step size of $0.013^{\circ}$ and time per step: 137.7~s. The sample $x=0.6$ was collected with a step size of $0.013^{\circ}$ and time per step: 90~s. Le Bail refinements were carried out in JANA2006~\cite{Petricek}.

\subsection{Thermodynamic measurements}

Dc and ac magnetic susceptibility and specific-heat measurements were conducted with a commercial Superconducting Quantum Interference Device (SQUID) magnetometer MPMS-XL and a Physical Property Measurement System (PPMS) by Quantum Design, respectively. For the specific-heat studies a heat-pulse relaxation technique was used. The
temperature- and field-dependent addenda were subtracted from the measured specific-heat values in the sample measurements. For each measurement of the dc magnetic susceptibility, the background signal of the sample holder was measured separately and subtracted from the total raw signal.

\section{Results}

\subsection{Growth optimization and structure determination}

\begin{figure}[h]
\begin{center}
\includegraphics[width=1.0\linewidth]{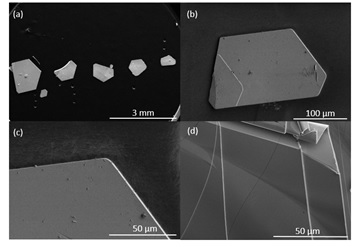}
\caption{SEM images of Ru$_{1-x}$Rh$_x$Cl$_3$ crystals: polycrystalline powder (a-b), layered surface morphology (c--d).}
    \label{tab:crystals} 
\end{center}
\end{figure}

Ru$_{1-x}$Rh$_x$Cl$_3$ crystals were obtained on various size scales: from micron sized suitable for X-ray diffraction to large foil-like platelets of several mm$^2$ appropriate for the magnetization and specific heat studies~(Fig.~\ref{tab:crystals}). 

In general, starting from the elements yielded polycrystalline powders of small platelet-like, black, partly intergrown crystals. The products form in that part of the ampule, where the starting materials were put so that we assume a gas phase assisted re-crystallization or a short-range transport for their formation. Only once a large crystal was obtained from the elements (for the nominal $x = 0.1$, see Table~\ref{tab:composition}).

\begin{table}[h]
    \caption{Experimentally determined compositions, $x^*$ (Rh/Rh+Ru from the EDS) and mass $m$ for the large crystals used in the magnetization studies (section~B). Crystals outlined in bold likely grew by chemical transport (see text). The crystal in italics was synthesized from the elemental precursors (see text).}
    \label{tab:composition}
        \centering
    \begin{tabular} {|c|c|c|c|} 
        \hline
        {\begin{tabular} {c}  Nominal \\ composition \end{tabular}}  &  {\begin{tabular} {c}  Average composition\\  (EDS, $K$-edge lines)\end{tabular}} & {\begin{tabular} {c} $\frac{1 - x^*}{x^*}$ \end{tabular}} & {\begin{tabular} {c} $m$, \\ mg \end{tabular}}
        \\
        \hline \hline
        Ru$_{0.9}$Rh$_{0.1}$Cl$_3$  &  \textbf{Ru$_{ 0.91(1)}$Rh$_{0.03(2)}$Cl$_{3.06(1)}$} & 0.97:0.03 & 2.12 \\
        \cline{2-4}
     & \textbf{Ru$_{ 0.95(2)}$Rh$_{0.07(3)}$Cl$_{2.97(2)}$} & {0.93:0.07} & 0.44 \\
        \cline{2-4}
        &  {Ru$_{0.86(2)}$Rh$_{0.10(1)}$Cl$_{3.04(2)}$} & {0.90:0.10} & 0.76 \\
        \cline{2-4}
        & Ru$_{ 0.86(1)}$Rh$_{0.10(1)}$Cl$_{3.03(1)}$ & {0.90:0.10} & 0.43 \\
        \cline{2-4}

        & Ru$_{ 0.83(1)}$Rh$_{0.10(1)}$Cl$_{3.07(1)}$ & 0.89:0.11 & 2.75 \\
        \cline{2-4}
        
        & {\it Ru$_{ 0.84(1)}$Rh$_{0.23(4)}$Cl$_{2.94(4)}$} & 0.79:0.21 & 0.44 \\
        \hline
        
        Ru$_{0.85}$Rh$_{0.15}$Cl$_3$  &  {Ru$_{ 0.89(3)}$Rh$_{0.18(2)}$Cl$_{2.93(3)}$} & 0.83:0.17 & 0.47 \\
        \hline
        Ru$_{0.8}$Rh$_{0.2}$Cl$_3$  & \textbf{Ru$_{ 1.01(1)}$Rh$_{0.02(1)}$Cl$_{2.98(1)}$} & 0.98:0.02 & 1.71  \\     \cline{2-4}
        & \textbf{Ru$_{ 0.88(1)}$Rh$_{0.07(1)}$Cl$_{3.05(1)}$} & 0.93:0.07 & 1.94 \\
        \cline{2-4}
        
        &Ru$_{0.79(1)}$Rh$_{0.20(1)}$Cl$_{3.01(1)}$ & 0.80:0.20 & 1.25 \\
        \cline{2-4}
        
        & Ru$_{ 0.80(2)}$Rh$_{0.21(3)}$Cl$_{2.98(2)}$ & 0.79:0.21 & 0.32 \\
        \hline
        
        Ru$_{0.7}$Rh$_{0.3}$Cl$_3$  & Ru$_{ 0.66(5)}$Rh$_{0.42(6)}$Cl$_{2.92(2)}$ & 0.61:0.39 & 0.3 \\
        \hline
        Ru$_{0.6}$Rh$_{0.4}$Cl$_3$  & Ru$_{ 0.74(1)}$Rh$_{0.29(2)}$Cl$_{2.98(2)}$ & 0.72:0.28 & 0.12 \\
        \cline{2-4}
        &  {Ru$_{ 0.68(2)}$Rh$_{0.45(2)}$Cl$_{2.88(1)}$} & 0.60:0.40 & 0.24 \\
       
        \hline
    \end{tabular}
\end{table}

Metal-halide precursors, in general, resulted in several large crystals, often intergrown or spreading from a common origin like flower petals. These crystals were always found at the same place where the precursors had been placed in the ampule, next to a ``powder'' of much smaller crystals, suggesting the same growth mechanism as above. Multiple attempts to control nucleation and encourage the growth of only large crystals by varying the tempering protocol (gradient, tempering times, cooling rate), ampule volume and gas partial pressures did not allow us to establish exact guidelines for tailored growth of large platelets. Typically, the higher gas-phase pressure encouraged a stronger nucleation and the formation of many small crystals.

When ampules with a larger volume of 28~cm$^3$ were used, large crystals sometimes formed at the initially empty side of the ampule, suggesting crystal growth via chemical vapor transport. However, these transported crystals had a very low Rh-content of $x < 0.1$ as per EDS (Table~\ref{tab:composition}), or were even pure $\alpha$-RuCl$_3$. The formation and transport properties of $\alpha$-RuCl$_3$ and RhCl$_3$ obviously diverge notably, imposing a substantial limitation on the attempts to synthesize Ru$_{1-x}$Rh$_x$Cl$_3$ by conventional chemical co-transport.

\begin{figure}[h]
\begin{center}
 \includegraphics[width=1\linewidth]{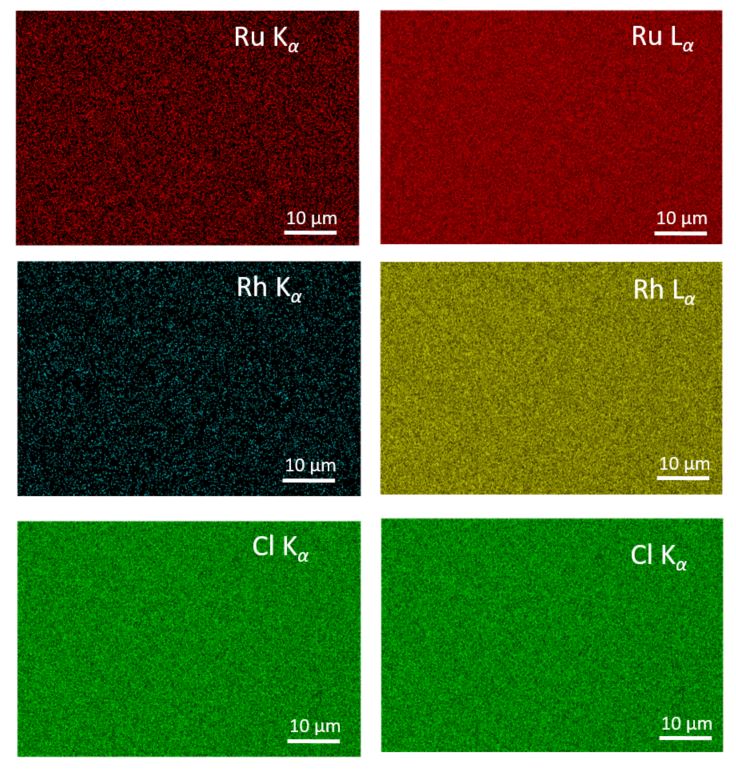}
\caption{EDS element mapping acquired at the $K$- (left column) and $L$-lines (right column) of a representative Ru$_{ 0.89(2)}$Rh$_{0.14(2)}$Cl$_{2.97(2)}$ crystal. Acquisition time: 24 minutes; Spectrum area: 26,863,239 counts.}
    \label{edxmaps} 

\end{center}
\end{figure}

In such multi-parameter growth conditions, the degree and uniformity of doping has to be addressed very carefully. In this spirit, all structural and magnetic studies were accompanied by EDS characterization. In order to determine the Rh doping with high accuracy, we implemented standards into EDS signal quantification (see~Experimental section~B). EDS mapping of the selected crystals confirmed uniform element distribution over large sample areas (Fig.~\ref{edxmaps}). Semi-qualitative EDS without reference compounds was quantified from the metal $L$-lines and showed a tendency to underestimate the Rh content significantly. The results of calibrated EDS were typically closer to the nominal compositions (cf. Table~\ref{tab:composition}). One unsolved issue is a deviation of the metal-to-chlorine content from 1:3. Since X-ray diffraction (see below) did not find indications for the presence of metal interstices or, alternatively, vacancies in the metal or chlorine substructures, we attribute these discrepancies to the measurement inaccuracy, that may arise e.~g.\ from surface roughness of the crystals which could not be polished. In order to study the magnetic response as a function of doping, we implemented the following renormalization to compare the specimens in a reliable way: $x^*$ was introduced as a portion of Rh related to the total metal content as determined by EDS (the last but one column in Table~\ref{tab:composition}).

Rhodium in the crystals with $x < 0.1$ (outlined in bold in Table~\ref{tab:composition}) could not be verified by EDS alone, because this technique does not deliver satisfactory accuracy at such low doping rates. The ultimate decision that the respective samples are in fact Rh-doped was made based on the magnetization data (see section~III.B). Large crystals from the batches with the nominal $x > 0.4$, on the other hand, were found to be inhomogeneously doped (Fig.~S1); these samples were not considered for magnetometry studies.

Crystal-structure elucidation was conducted by X-ray diffraction on small suitable crystals and powders of Ru$_{1-x}$Rh$_x$Cl$_3$ ($x = 0 - 0.6$) and RhCl$_3$. We refined the latter because the earlier published crystallographic data~\cite{barnighausen1964} did not suffice for comparison.

\begin{figure}[h]
\begin{center}
 \includegraphics[width=1.0\linewidth]{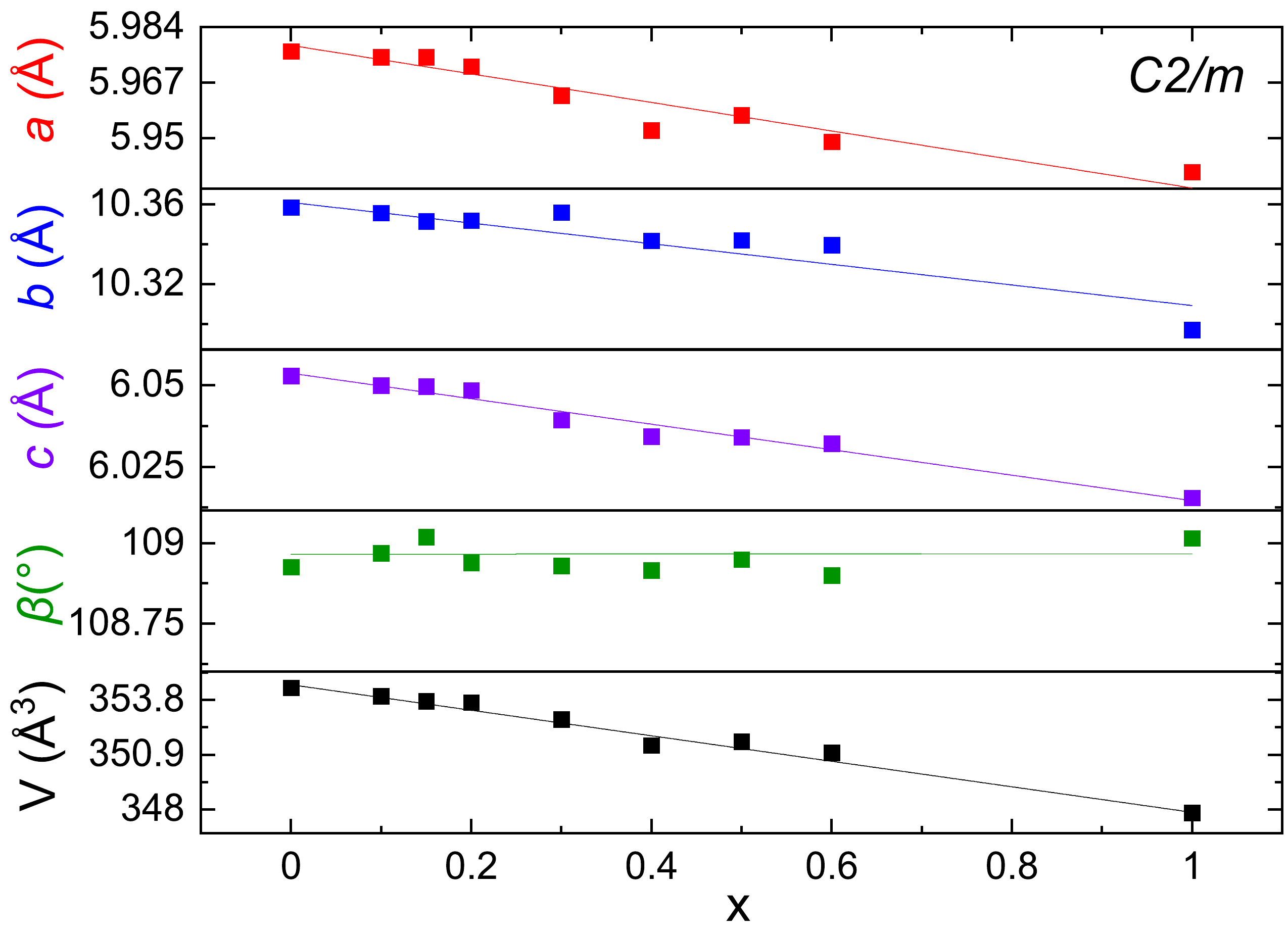}
\caption{The refined unit cell parameters of Ru$_{1-x}$Rh$_x$Cl$_3$ as a function of the nominal Rh content $x$ from the PXRD data collected at room temperature on ground powders.}
    \label{pxrd} 

\end{center}
\end{figure}

Owing to the platelet-like crystal morphology, the PXRD patterns of ``polycrystalline'' powders showed very prominent texture effects, namely significant intensity enhancement of the $(00l)$ reflections and notable weakening of the $(hk0)$ reflections. Attempts to minimize the preferred orientation of the crystallites by mixing them with amorphous silica and performing X-ray experiments in capillaries did not solve the issue. The texture could be partly mitigated by grinding, but this, in turn, introduced noticeable peak broadening in all samples. Both binary and ternary chlorides exhibited similar peak widths. Using the Scherrer equation, we estimate the crystallite size in an as-synthesized sample as 84~nm, and in a ground sample as 56~nm. Alongside the decreasing particle size, the formation of stacking faults by the mechanical impact could also contribute to the peak broadening. We did not observe any peak splitting in the PXRD patterns or any other indications for multi-phase samples with varying $x$ in our specimen.
For these reasons, we performed only a Le Bail fitting of the PXRD data of the ground samples (Fig.~S2, Table~S1 in~SI). The full pattern quality did not suffice for Rietveld refinements. The computed unit cell parameters of Ru$_{1-x}$Rh$_x$Cl$_3$ ($x = 0-1$) in the monoclinic lattice are plotted in Fig.~\ref{pxrd} against the nominal Rh content $x$.

The resultant lattice parameters monotonously decrease as a function of $x$ in accordance with Vegard's law for solid-state solutions. Due to the strongly suppressed intensity of the $(hk0)$ reflections (cf.~Fig.~S3) in some Ru$_{1-x}$Rh$_x$Cl$_3$ specimen, the parameters $b$ and $\beta$ were less accurately determined and, thus, show a somewhat larger spread of values around the respective trendline (Fig.~\ref{pxrd}).
Furthermore, we attempted to index the PXRD patterns in the alternative space groups proposed for $\alpha$-RuCl$_3$. Indexing in the rhombohedral $R\bar{3}$ space group was incompatible with the experimentally observed reflection conditions. Indexing in the trigonal group $P3_{1}12$ (which was considered in previous works ~\cite{Ziatdinov, Lampen-Kelley2017}) gives reasonable $R$-values (Table~S2) and linear dependencies of the lattice parameters on doping (Fig.~S5), but is not confirmed by any subsequent single-crystal refinements (see below). As it is shown from our indexing, the powder pattern seem to be compatible with both space groups due to high texture effects and peak profiles broadening. However, SCXRD data confirm $C$2/$m$ to be the correct one for the Ru$_{1-x}$Rh$_x$Cl$_3$ series with  $ 0 \le x \le 1$.

\begin{figure}[h]
\begin{center}
 \includegraphics[width=1.0\linewidth]{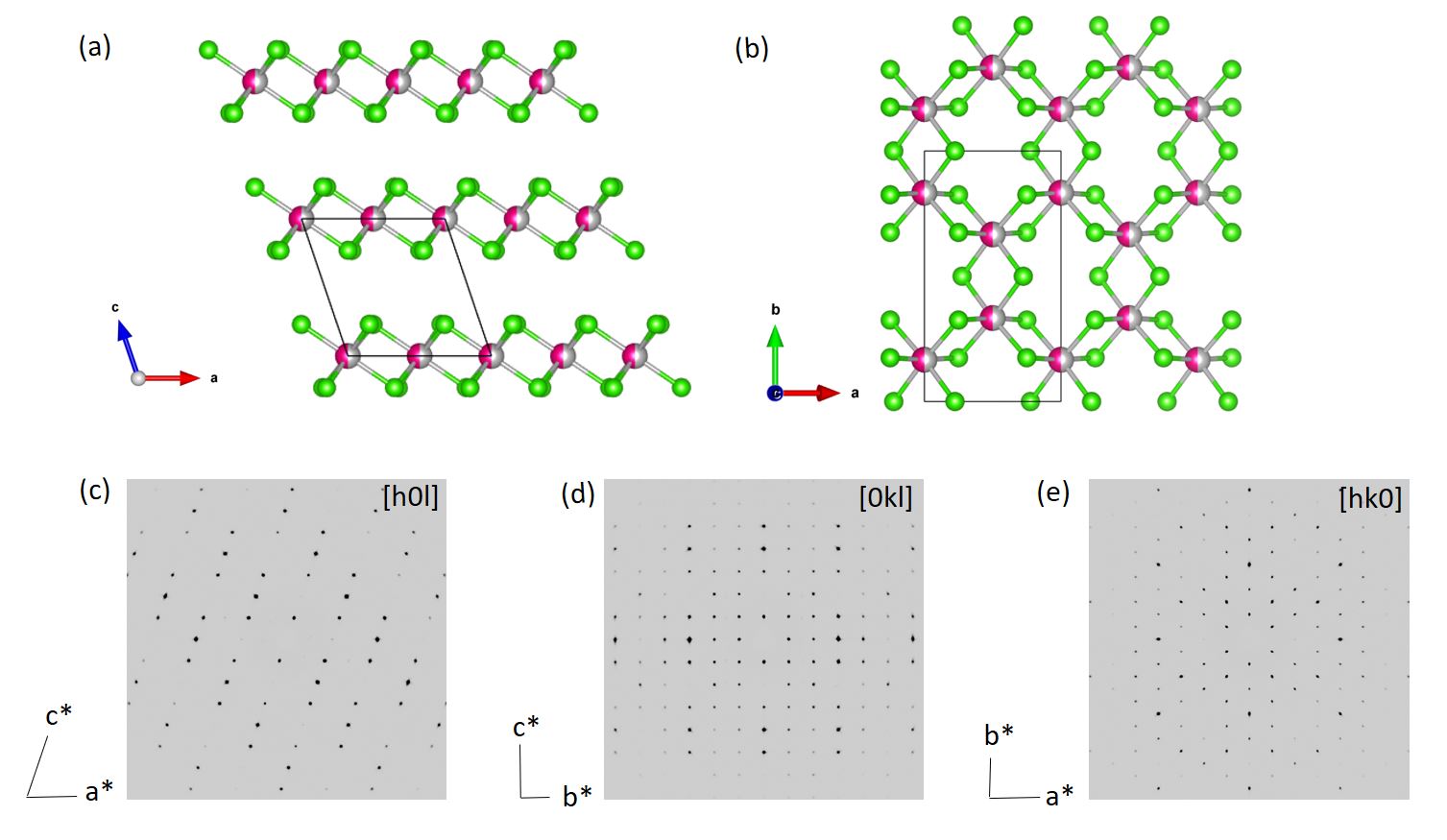}
\caption{(a,b) Crystal structure of Ru$_{1-x}$Rh$_x$Cl$_3$ plotted with the VESTA visualization program~\cite{Momma:db5098}. Ru -- red, Rh -- grey, Cl -- green. (c--e) Reciprocal-space layers reconstructed from a SCXRD dataset for Ru$_{0.78(4)}$Rh$_{0.24(4)}$Cl$_{2.98(4)}$ measured at 100~K.}
    \label{tab:structure_image} 

\end{center}
\end{figure}

The results of the SCXRD measurements and refinements are summarized in Tables~\ref{tab:structure_single_crystals}, ~\ref{tab:structure_single_crystals_low}, SIII and SIV. All studied specimens irrespective of the doping level were found to be isostructural to the parent monoclinic structures of $\alpha$-RuCl$_3$ and RhCl$_3$ (sp.\ gr.\ $C2/m$), Fig.~\ref{tab:structure_image} a,b. The reciprocal-space layers (Fig.~\ref{tab:structure_image} c-e) reconstructed from an SCXRD dataset of an exemplary Ru$_{0.78(4)}$Rh$_{0.24(4)}$Cl$_{2.98(4)}$ crystal demonstrate faint diffuse scattering along the $c^\ast$ direction, indicating a low density of stacking faults. 
As mentioned above, reflection conditions and refinement results are not compatible with higher symmetric space groups. 
Atom positions and anisotropic displacement parameters for all samples are listed in Tables SIII and~SIV. Low-temperature single crystal diffraction experiments performed for selected compositions at $T = 100$~K showed no evidence of structural modifications like the formation of an ordered superstructure. However, diffuse streaks seem more intense and residual electron density peaks in the difference Fourier maps ($F_o-F_c$) are larger for the low-temperature data of all investigated compounds, which may be taken as evidence for an increased amount of stacking faults. Albeit, no hints for a change of the lattice symmetry indicative of a structural phase transition as discussed for $\alpha$-RuCl$_3$ was observed. 
Clear evidence for a structural change at least for samples with low doping level $x \lesssim 0.1$, however, comes from magnetization data (see below). This apparent contradiction may be explained by a suppressed (by rapid quenching, e.g.) or an incomplete phase transition. The structural change from an {\it ABC} based stacking (in space group $C$2/$m$) to an {\it AB} based stacking (space group $R\bar{3}$) should proceed via an  increasing number of stacking faults, giving rise to the emergence of different ordering temperatures $T_N$ in magnetization and to diffuse scattering intensities in diffraction experiments. The overall symmetry of the diffraction pattern will, however, not change until the phase transition is complete or nearly complete. For some of the investigated single crystals, diffuse scattering along the $c^\ast$-direction was indeed found to be more prominent, pointing towards a noticeable amount of stacking faults. Although diffuse streaks were not as severe as reported for $\alpha$-RuCl$_3$ \cite{Johnson2015} or Ru$_{1-x}$Cr$_x$Cl$_3$ \cite{Roslova2019} previously, the same typical pattern was observed: alternating rows in ${[0kl]}$ with discrete intense Bragg spots for rows with $k=3n$ ($n$ -- integer) and diffuse scattering intensities between Bragg spots for $k=3n+1$ and $k=3n+2$ (Fig.~S6b).  Bragg spots often were a little bit smeared out in the $l$-direction. Additionally, diffuse intensity stripes were also visible in the $[h0l]$ reciprocal layers (Fig.~S6a), but not in the reciprocal layers $[hk0]$ (Fig.~S6c, f). Interestingly, the crystal of binary RhCl$_3$ showed agglomeration of scattering intensity around non-integer positions with $l=n+1/3$ ($n$ -- integer) (Fig.~S6d, e). 
The course of the SCXRD-derived unit cell parameters and cell volumes with renormalized $x^*$ is given in Fig.~S4. Due to the different measuring conditions and instruments, the refined unit cell parameters scatter visibly, but they indicate the same trend as the one observed from the PXRD data. The unit cell volumes derived from both methods on the other hand match quite well. From the scattering experiments we conclude that rhodium appears to be statistically distributed over the metal atom sites in the $\alpha$-RuCl$_3$-type structure for all studied doping rates $x$.

\begin{table*}[p]
    \centering
    \resizebox{\textwidth}{!}{
    \begin{tabular}{c|c|c|c|c|c}
        \hline
        Nominal & Ru$_{0.9}$Rh$_{0.1}$Cl$_3$ &
        Ru$_{0.8}$Rh$_{0.2}$Cl$_3$ &
        Ru$_{0.7}$Rh$_{0.3}$Cl$_3$  &
        Ru$_{0.5}$Rh$_{0.5}$Cl$_3$ &
        RhCl$_3$
        \\
        composition & & & & & \\
        \hline
     EDS & Ru$_{0.76(2)}$Rh$_{0.29(2)}$Cl$_{2.96(1)}$  &
     Ru$_{0.78(4)}$Rh$_{0.24(4)}$Cl$_{2.98(4)}$ &
     Ru$_{0.56(5)}$Rh$_{0.46(4)}$Cl$_{2.97(3)}$ &
     Ru$_{0.60(2)}$Rh$_{0.48(1)}$Cl$_{2.92(1)}$ &
     RhCl$_3$ \\
       \hline
   $1-x*:x*$ & 0.73:0.27 & 0.77:0.23 & 0.55:0.45 & 0.56:0.44 & 1 \\
       \hline
   Refined & Ru$_{0.7}$Rh$_{0.3}$Cl$_3$  &  Ru$_{0.8}$Rh$_{0.2}$Cl$_3$ &
    Ru$_{0.6}$Rh$_{0.4}$Cl$_3$  &
    Ru$_{0.6}$Rh$_{0.4}$Cl$_3$ &
    RhCl$_3$
     \\
        composition & & & & & 
     \\
     \hline
   Crystal system & \multicolumn{5} {c} {Monoclinic, ${C2/m}$, 4} 
    \\
 \hline
   Wavelength & \multicolumn{5} {c} {Mo K$_\alpha$, 0.71073 \AA   } 
    \\
   \hline
   Temperature & \multicolumn{5} {c} {298 K} 
    \\
   \hline
   Range of collection & $3.56 \degree \leq \theta  \leq 29.08 \degree$ & $3.564 \degree \leq \theta  \leq 29.095 \degree$ & $3.573 \degree  \leq \theta  \leq 29.08 \degree$ & $3.57 \degree \leq \theta \leq 29.108 \degree$ & $ 3.588 \degree \leq \theta \leq 29.078 \degree$ \\ 
        \hline
    Index ranges &
    -8 $ \leq$ h $ \leq$ 8  & -8 $ \leq$ h $ \leq$ 8 & -8 $ \leq$ h $ \leq$ 7 & -8 $ \leq$ h $ \leq$ 7 & -8 $ \leq$ h $ \leq$ 8 \\
      & -12 $ \leq$ k $ \leq$ 14 & 
     -14 $ \leq$ k $ \leq$ 14 & -14 $ \leq$ k $ \leq$ 14 & -14 $ \leq$ k $ \leq$ 14 & -14 $ \leq$ k $ \leq$ 13 \\
     
    &
    -8 $ \leq$ l $ \leq$ 8  & -8 $ \leq$ l $ \leq$ 7 & -8 $ \leq$ l $ \leq$ 8 & -8 $ \leq$ l $ \leq$ 8 & -8 $ \leq$ l $ \leq$ 8 \\  \hline
    Number of reflections & 2024 & 2254 & 1995 & 2037 & 1851 \\
         \hline
  
   Absorption coefficient $\mu$,  & 6.544 & 6.517 & 6.626 & 6.614 & 6.947 \\ 
    
 mm$^{-1}$ & & & & &
    \\ 
        \hline
    Crystal density $\rho$,  & 3.911 & 3.909 & 3.939 & 3.932 & 4.008 \\ 
    
 g $\cdot$ cm$^{-3}$ & & & & &
    \\
    \hline
   Structure refinement & \multicolumn{5} {c} {Full-matrix least squares based on $F_{0}^{2}$} \\ 
   \hline
  Data/ & 508/22 & 508/22 & 503/22 & 507/22 & 494/22 \\ 
  parameters & & & & & \\
   \hline
   $R_{int}$ & 0.0432 & 0.0345 & 0.0562 & 0.0555 & 0.0442 \\ 
        \hline
   $R_{I}$ / $wR_2$ (all) & 0.019/0.0421 & 0.0196/0.0461 & 0.0281/0.06 & 0.0345/0.0694 & 0.0311/0.0724\\
    \hline
 GoF (all) & 1.015 & 1.068 & 1.023 & 1.006 & 1.181\\
             \hline
Lattice & $a = 5.978(1)$ & $a = 5.976(1)$ & $a = 5.963(2)$ & $a = 5.969(2)$ & $a = 5.936(1)$ \\
parameters \AA & $b = 10.334(2)$ & $b = 10.339(2)$ & $b = 10.321(3)$ & $b = 10.326(3)$ & $b = 10.290(3)$\\
& $c = 6.035(1)$ & $c = 6.035(1)$ & $c = 6.020(2)$ & $c = 6.023(2)$ & $c = 5.992(2)$ \\
& $\beta = 108.68(2)$ & $\beta = 108.71(2)$ & $\beta = 108.68(2)$ & $\beta = 108.70(2)$ & $\beta = 108.63(2)$ \\
           \hline
Cell volume, \AA$^3 $ & 353.2(1) & 353.1(1) & 351.0(2) & 351.6(2) & 346.8(2)\\
\hline
Residual electron & $+1.0/-1.2$ & $+1.2/-1.1$ & $+0.8/-1.7$ & $+2.3/-1.4$ & $+5.6/-1.6$ \\
density, $e$ \AA $^{-3}$ & & & & & \\
\hline

    \end{tabular}}
    
    \caption{Crystallographic data for Ru$_{1-x}$Rh$_x$Cl$_3$ single crystals collected at 298~K. All compounds crystallize in a monoclinic unit cell (sp.\ gr.\ $C2/m$ (No.~12), $Z = 4$). Nominal compositions of the batches, from which the crystals were picked out, are given alongside the experimentally determined compositions of these crystals by calibrated EDS.}
    \label{tab:structure_single_crystals}
\end{table*}

\begin{table*}[h]
    \centering
    \begin{tabular}{c|c|c|c}
        \hline
        Nominal & Ru$_{0.9}$Rh$_{0.1}$Cl$_3$ &
        Ru$_{0.8}$Rh$_{0.2}$Cl$_3$  &
        Ru$_{0.6}$Rh$_{0.4}$Cl$_3$ 
        \\
        composition & & & \\
        \hline
        
     EDS & Ru$_{0.76(2)}$Rh$_{0.18(2)}$Cl$_{3.06(3)}$   &
     Ru$_{0.78(4)}$Rh$_{0.24(4)}$Cl$_{2.98(4)}$ &
     {Ru$_{0.69(4)}$Rh$_{0.46(7)}$Cl$_{2.85(5)}$} \\
       \hline
   $1-x*:x*$ & 0.81:0.19 & 0.77:0.23 & 0.6:0.4  \\
       \hline
   Refined & Ru$_{0.8}$Rh$_{0.2}$Cl$_3$   &
    Ru$_{0.8}$Rh$_{0.2}$Cl$_3$ &  Ru$_{0.6}$Rh$_{0.4}$Cl$_3$
    \\
        composition & & &
     \\
     \hline
   Crystal system & \multicolumn{3} {c} {Monoclinic, ${C2/m}$, 4} 
    \\
 \hline
   Wavelength & \multicolumn{3} {c} {Mo K$_\alpha$, 0.71073 \AA   } 
    \\
   \hline
   Temperature & \multicolumn{3} {c} {100 K} 
    \\
   \hline
   Range of collection & $3.580 \degree  \leq \theta  \leq 27.715 \degree$  & $3.573 \degree  \leq \theta \leq 35.88 \degree$ & $3.592 \degree \leq \theta \leq 44.83 \degree$ \\ 
        \hline
    Index ranges &
    -9 $ \leq$ h $ \leq$ 10  & -9 $  \leq$ h $ \leq$   9 & -11 $ \leq$ h $ \leq$ 11 \\
      & -17 $ \leq$ k $ \leq$ 7 & -16 $ \leq$ k $ \leq$ 16 & 
     -11 $ \leq$ k $ \leq$ 11  \\
     
    &
    -10 $ \leq$ l $ \leq$ 10  & -9 $ \leq$ l $ \leq$  9 & -20 $ \leq$ l $ \leq$ 20 \\  \hline
  Number of reflections  & 10405 & 9311  & 16861 \\ 
         \hline
 
   Absorption coefficient $\mu$,  & 6.550  & 6.523 & 6.673  \\ 
    
 mm$^{-1}$ & & & 
    \\ 
        \hline
    Crystal density $\rho$,  & 3.937  & 3.931 & 3.967  \\ 
    
 g $\cdot$ cm$^{-3}$ & & & 
    \\
    \hline
    
   Structure refinement & \multicolumn{3} {c} {Full-matrix least squares based on $F_{0}^{2}$} \\ 
   \hline
  Data/ & 956/22 & 857/22 & 1469/22   \\ 
  parameters & & &  \\
  
             \hline
   $R_{int}$ & 0.0453  & 0.0353 & 0.0279 \\
        \hline
   $R_{I}$ / $wR_2$ & 0.0268/0.0639 & 0.0186/0.0492 & 0.0133/0.0333  \\
    \hline
   GoF & 1.189  & 1.170 & 1.165 \\
             \hline
Lattice & $a = 5.9633(2)$ & $a =5.9720(1) $ & $a = 5.9569(1)$   \\
parameters \AA & $b = 10.3236(3)$  & $b = 10.3359(1)$ & $b = 10.3114(1)$\\
& $c = 6.0090(3)$  & $c = 6.0065(1)$ & $c = 5.9877(1)$ \\
& $\beta = 108.748(5)$ & $\beta =108.754(2)$ & $\beta = 108.690(2)$   \\
           \hline
Cell volume, \AA$^3$ & 350.30(3) & 351.073(10) & 348.393(1)  \\
\hline
Residual electron & $+10.0/-1.4$ & $+2.7/-1.3$ & $+5.4/-1.4$ \\
density, e \AA$^{-1}$ & & &  \\
\hline
    \end{tabular}
    \caption{Crystallographic data for Ru$_{1-x}$Rh$_x$Cl$_3$ single crystals collected at 100~K. All compounds crystallize in a monoclinic unit cell (sp.\ gr.\ $C2/m$ (No.~12), $Z = 4$). Nominal compositions of the batches, from which the crystals were picked out, are given alongside the experimentally determined compositions of these crystals by calibrated EDS.}
    \label{tab:structure_single_crystals_low}
\end{table*}

\subsection{Magnetization measurements}

\begin{figure*}[t]
\begin{minipage}{0.49\linewidth}
\begin{center}
\includegraphics[width=1\linewidth]{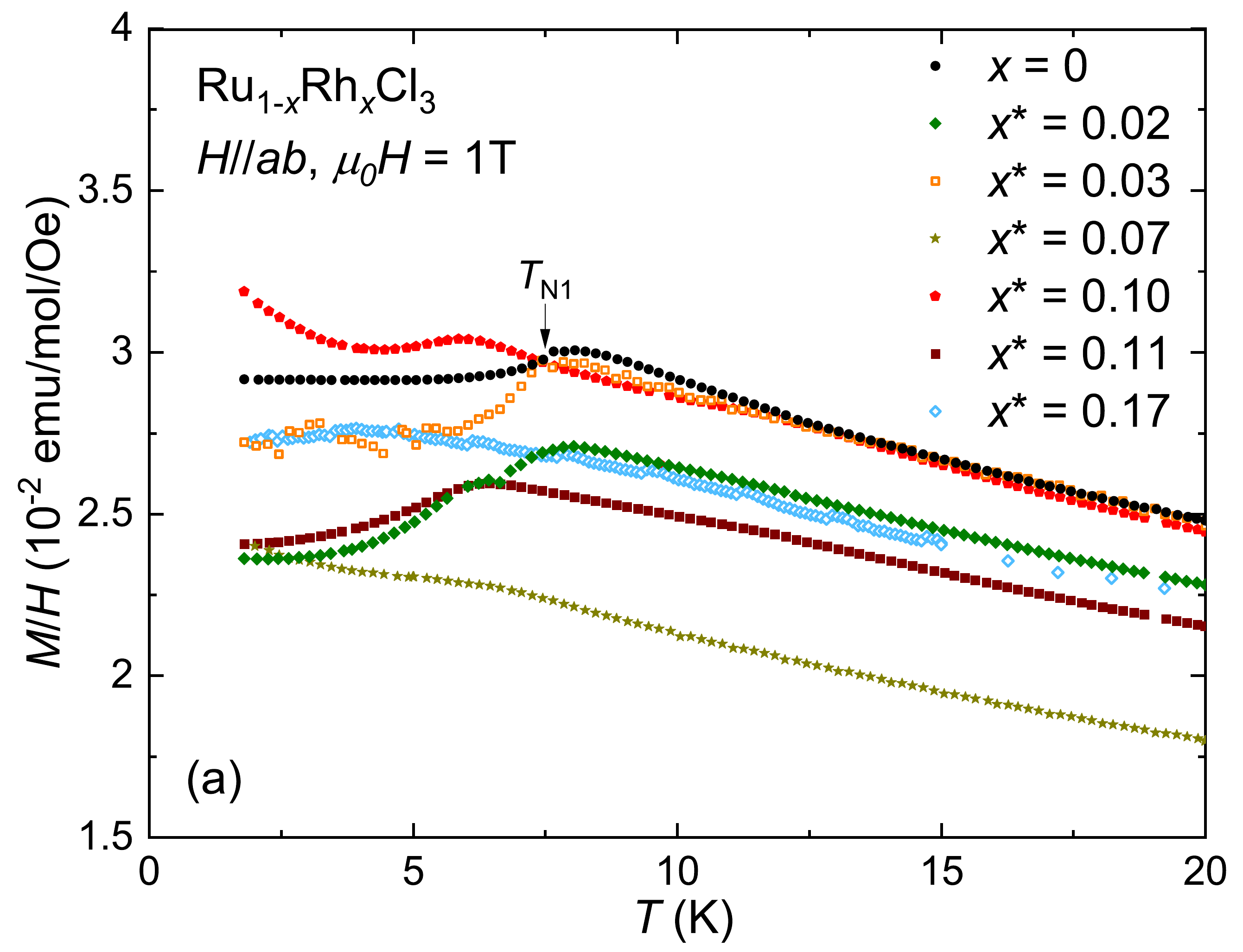}
\includegraphics[width=1\linewidth]{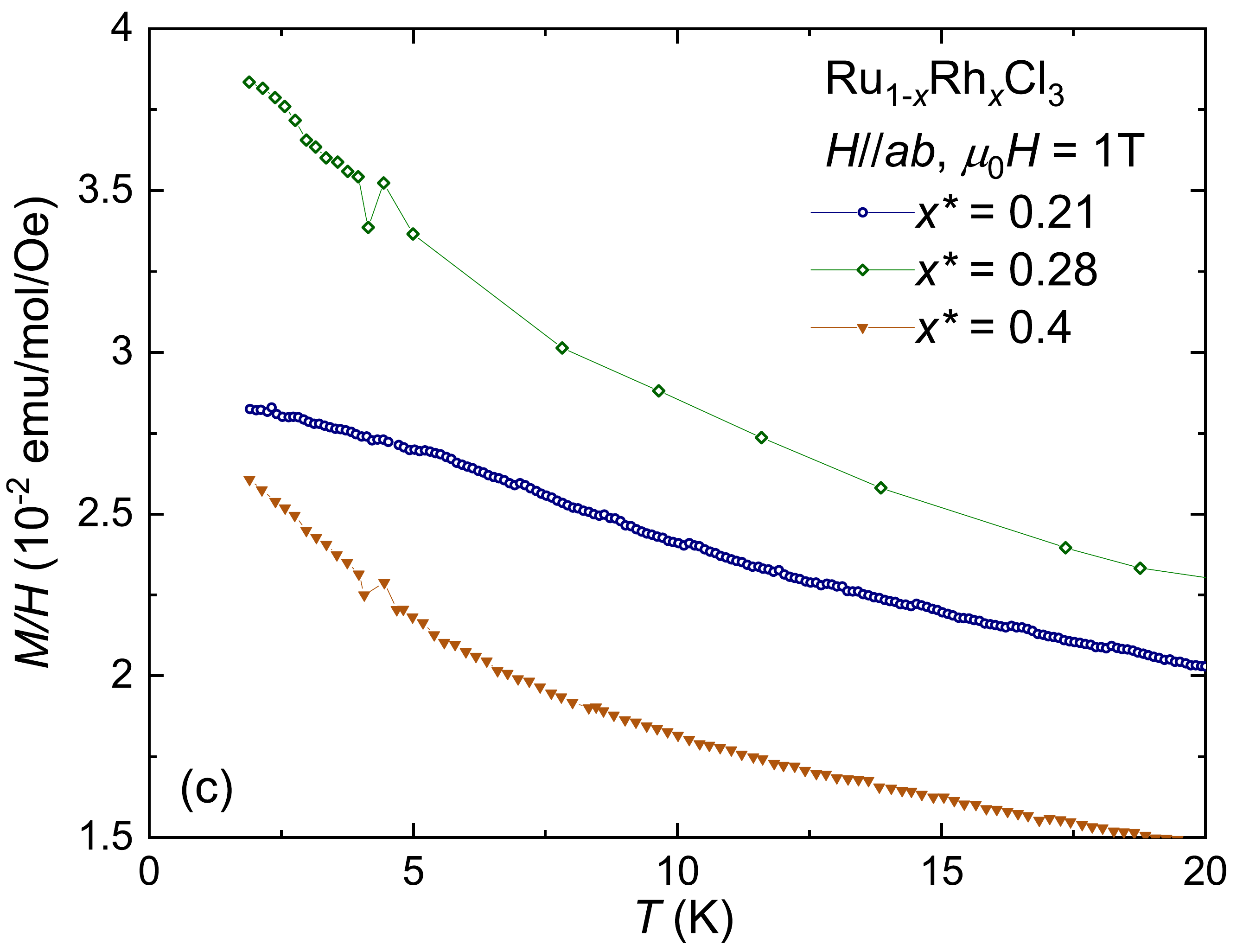}
\end{center}
\end{minipage}
\hfill
\begin{minipage}{0.49\linewidth}
\begin{center}
\includegraphics[width=1\linewidth]{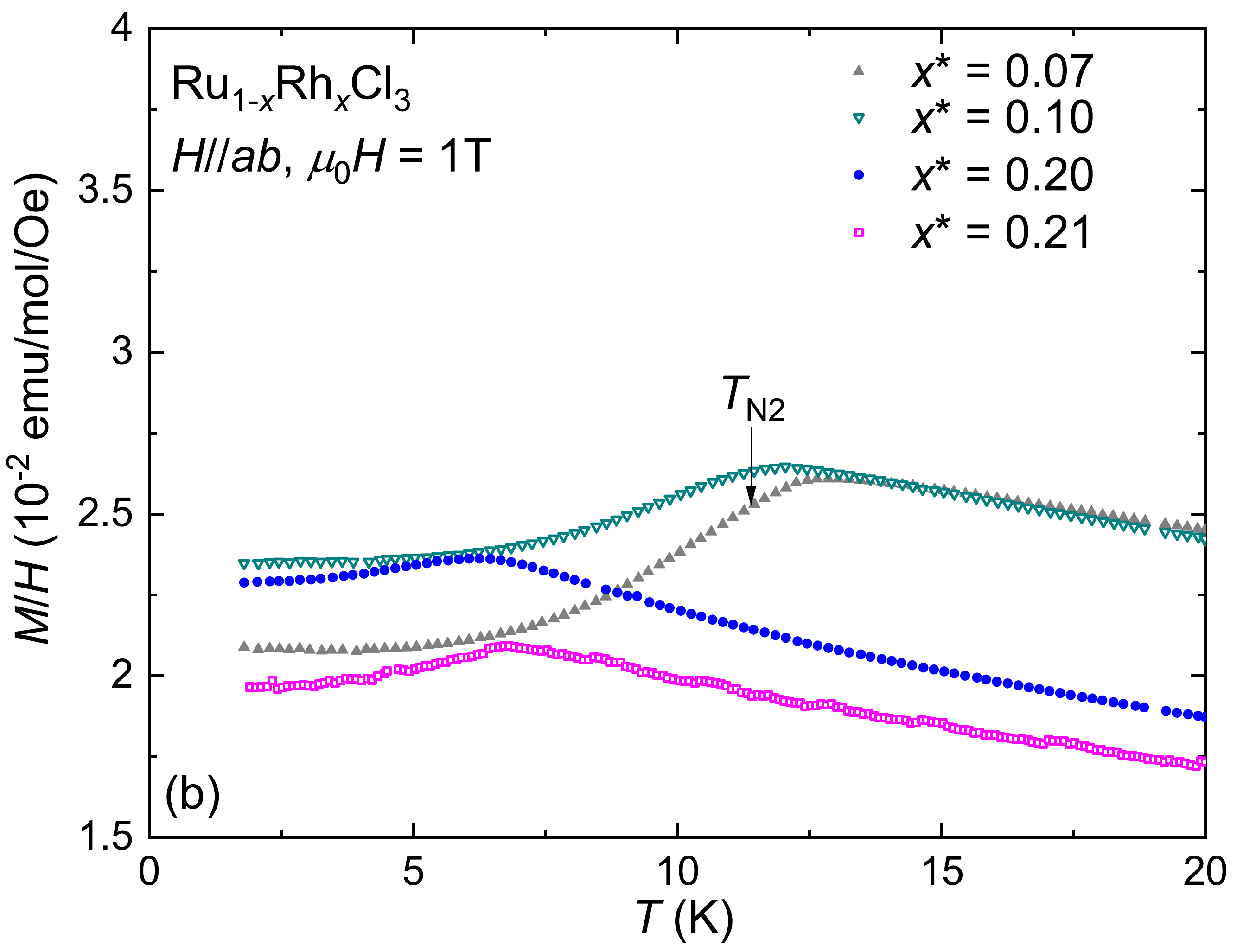}
\includegraphics[width=1\linewidth]{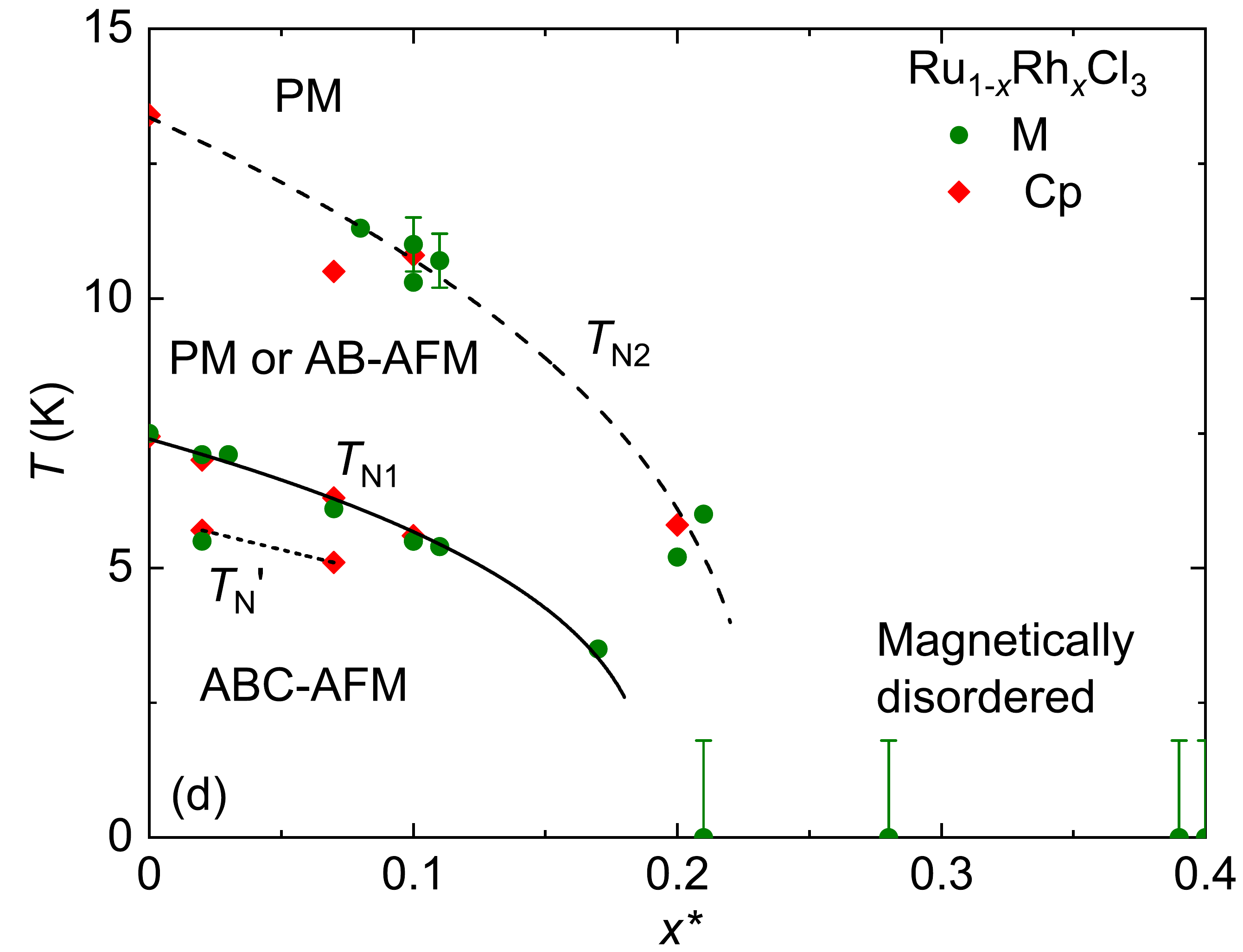}
\end{center}
\end{minipage}
\caption{(a) Normalized magnetization of Ru$_{1-x}$Rh$_x$Cl$_3$ crystals in a magnetic field in the $ab$ plane as a function of temperature. Only crystals assumed to harbor an $ABC$ stacking were selected for this figure. Successive thermal cycling through the structural transition was used to reduce the amount of stacking faults. (b) Same figure for crystals assumed to harbor an $AB$ stacking.
(c) Same figure for crystals with $x\geq 0.21$, whose stacking sequence cannot be determined unambiguously. (d) $(x*-T)$ phase diagram of Ru$_{1-x}$Rh$_x$Cl$_3$. The green circles and red diamonds are experimental points from magnetization and specific heat, respectively. Solid lines are guides to the eye and they show the evolution of the various N\'eel temperatures $T_{\mathrm N1}$, $T_{\mathrm N2}$ and $T_{\mathrm N}'$. Each crystal showing absence of magnetic order is represented by a point on the $T=0$ axis. The different antiferromagnetic (AFM) and paramagnetic (PM) phases are indicated.
}
\label{MvsT}
\end{figure*}

The temperature dependence of the normalized low-temperature magnetization of Ru$_{1-x}$Rh$_x$Cl$_3$ is represented in Fig.~\ref{MvsT} for a magnetic field applied within the $ab$ plane. While for $x^*\leq 0.21$ a maximum is observed in the magnetic susceptibility, indicative of magnetic order, samples with $x^* > 0.21$ do not show signs of magnetic order. The magnetic ordering temperature is defined at the temperature, where $d(MT)/dT$ undergoes its maximum. Interestingly, strong differences in the magnetic ordering temperature were observed between crystals with the same Rh content. A similar sample dependence was reported for $\alpha$-RuCl$_3$ and it was interpreted as the consequence of different stacking sequences~\cite{Cao2016}. The drawing of the temperature versus Rh content phase diagram $(x^*-T)$ in Fig.~\ref{MvsT}(d) enables us to sort the crystals with $x^*<0.2$ in two groups in analogy with the sample dependence observed in $\alpha$-RuCl$_3$~\cite{Cao2016} related to the stacking sequence. The ordering temperature $T_{\mathrm N1}$ of the crystal from the first group decreases from $T_{\mathrm N1}$=7.5\,K for pure $\alpha$-RuCl$_3$ ($x$=0) and vanishes around $x^*=0.2$. These crystals are expected to harbor the structure with $ABC$ stacking as represented in Fig.~\ref{tab:structure_image} (a). The second group of crystal orders at higher temperatures. Their ordering temperatures are situated on a second line in the phase diagram starting from $T_{\mathrm N1}$=13\,K for pure $\alpha$-RuCl$_3$ ($x$=0) and vanishing for a composition around $x^*=0.25$. In analogy with previous work on $\alpha$-RuCl$_3$~\cite{Cao2016} we propose, that these crystals exhibit a stacking sequence $AB$. This shows that the Ru$_{1-x}$Rh$_x$Cl$_3$ crystals grow with different stacking even within the same batch and that most samples are homogeneous in terms of stacking. A third transition $T_{\mathrm N}'$ occurs in the sample $x^*=0.02$ and it will be discussed later based on specific heat measurements.

\begin{figure*}[t]
\begin{center}
\includegraphics[width=0.9\linewidth]{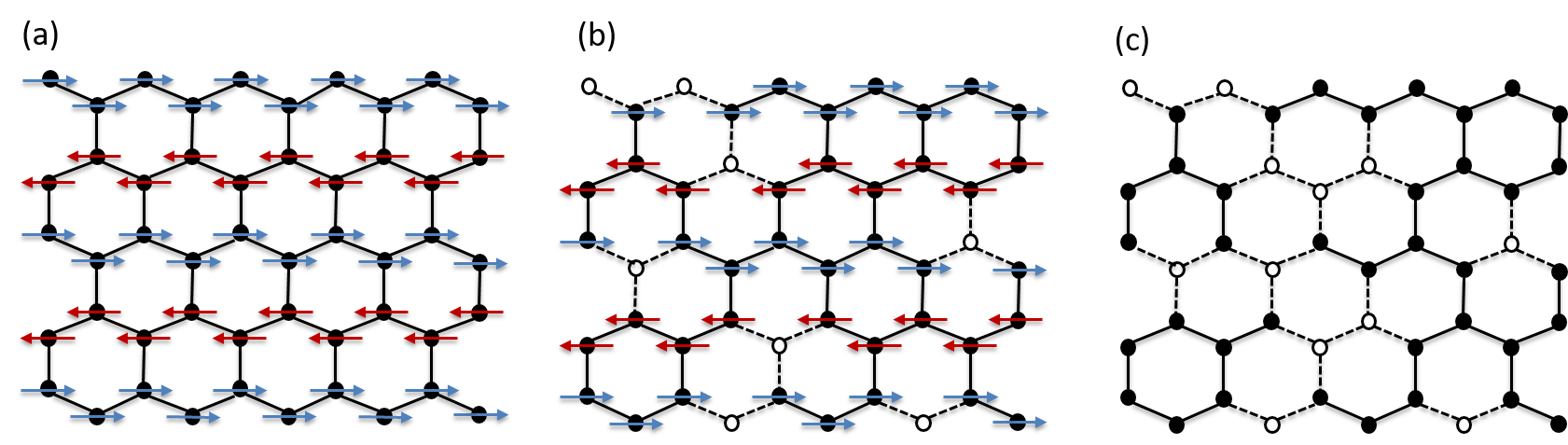}
\caption{(a) A schematic view of the magnetic ground state of the honeycomb  $\alpha$-RuCl$_3$ lattice called zigzag order~\cite{Sears2015}. (b) Honeycomb lattice with 0.16 vacancies per formula unit. The magnetic order subsists despite the large number of vacancies. A small tilting of the magnetic moments induced by neighboring vacancies is expected~\cite{Maryasin2013,Andrade2014}, it is not represented in the figure for simplicity. (c) Honeycomb lattice with 0.24 vacancies per formula units. The magnetic ground state is disordered and it is not frozen.} 
\label{honeycomb}
\end{center}
\end{figure*}

Based on these experimental results we draw a sketch of the magnetic lattice in our series for three different substitutions $x^*$: no vacancies, for 0.16 vacancies per formula unit and for 0.24 vacancies per formula unit (Fig.~\ref{honeycomb} (a-c)). While the zigzag magnetic order is preserved in the case $x^*$= 0.16, despite the dilution of the magnetic lattice, for $x^*$ = 0.24 the magnetic spins show no static long-range order anymore. This schematic view helps to visualize the robustness of the magnetic ground state of $\alpha$-RuCl$_3$ with respect to magnetic dilution.

\begin{figure}[t]
\begin{center}
\includegraphics[width=0.9\linewidth]{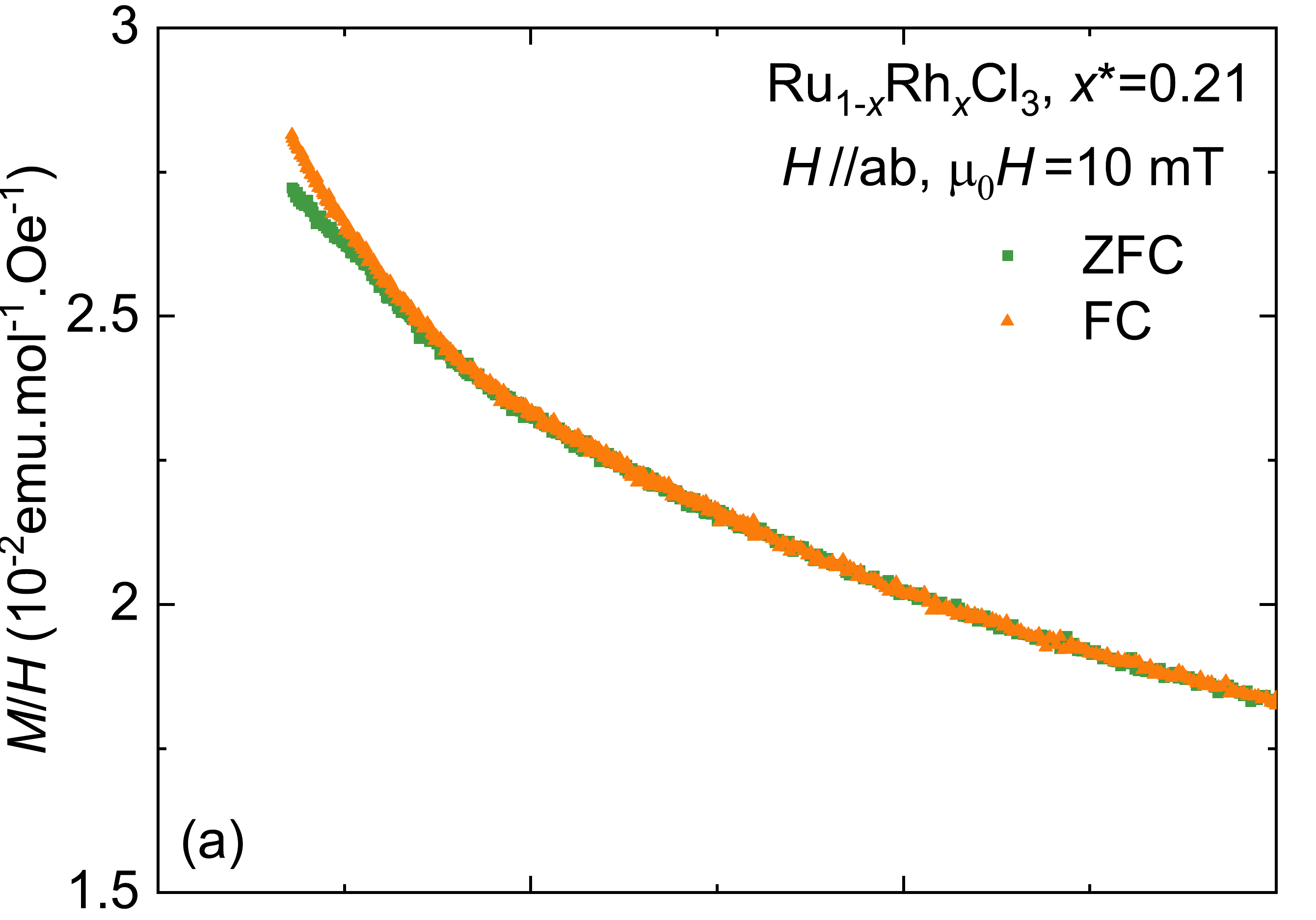}
\includegraphics[width=0.9\linewidth]{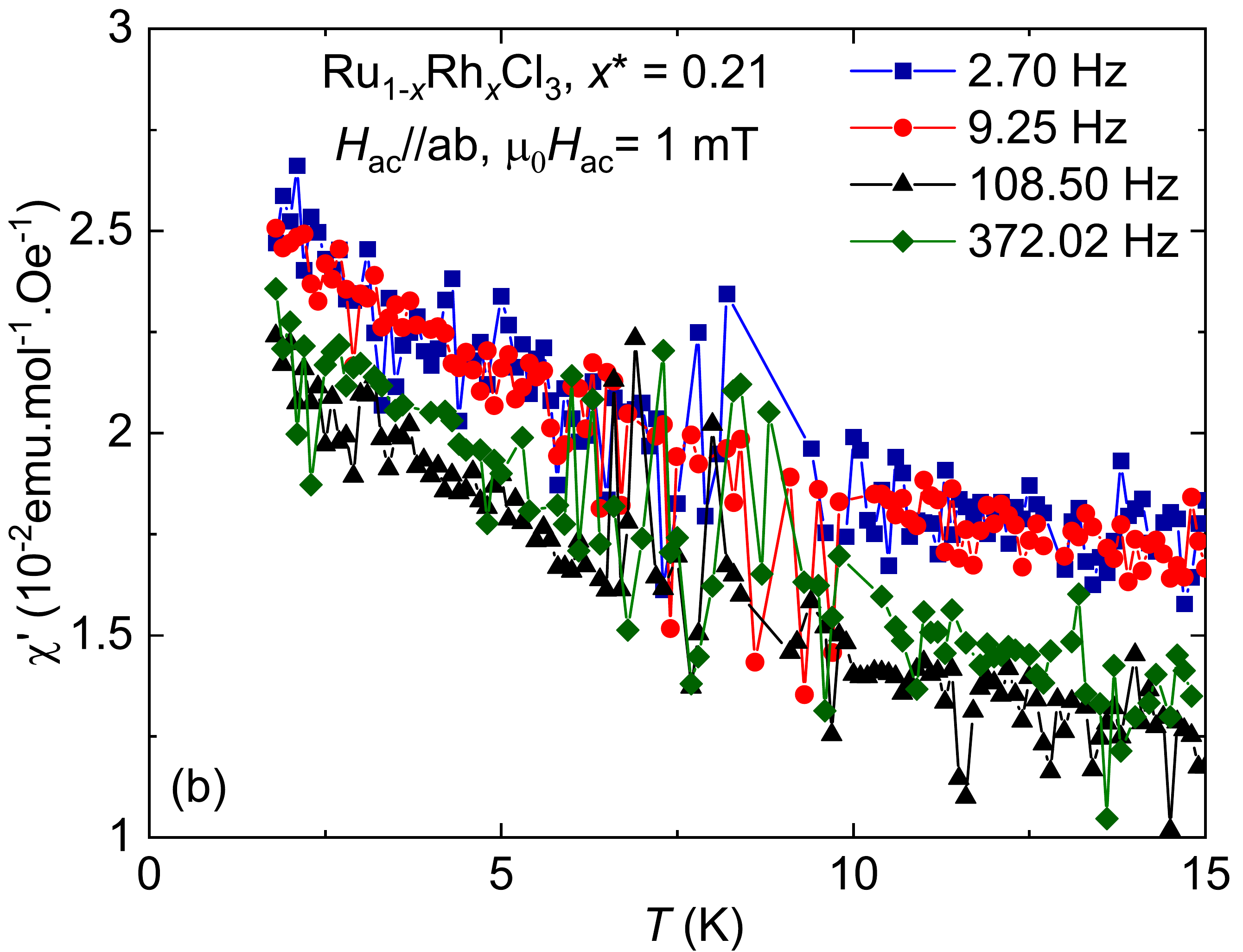}
\caption{(a)  Zero-field cooled (ZFC) and field-cooled (FC) normalized magnetization of Ru$_{1-x}$Rh$_x$Cl$_3$ for $x^*=0.21$ in a field of $\mu_0 H=0.01$ T applied in the $ab$ plane as a function of temperature. The sample was not warmed up above 15~K between the two measurements to avoid any unwanted evolution upon thermal cycling. A splitting between the ZFC and the FC curves can be observed at very low temperatures. (b)  Real part of the in-plane ac magnetic susceptibility of Ru$_{1-x}$Rh$_x$Cl$_3$ for $x^*=0.21$ as a function of temperature for different frequencies.} 
\label{ac}
\end{center}
\end{figure}

In order to search for signatures of an eventual spin-glass state for $x^* > 0.2$, the magnetization was measured in a low magnetic field of $\mu_0H=0.01$ T for $x^*$=0.21 both under zero-field cooled (ZFC) and field-cooled (FC) conditions (see Fig.~\ref{ac} (a)). A small splitting between the zero-field and field-cooled magnetization can be observed for $T \leq$ 3~K. It may indeed indicate a glassy behavior but it may also results from a small amount of magnetic impurities, which are often seen in magnetic measurements in low external fields. In order to test further the possibility of a spin-glass state, we performed ac magnetic susceptibility measurements as a function of temperature and frequency. The real part $\chi'$ is represented in Fig.~\ref{ac} (b). The imaginary part $\chi''$ remained around zero within the experimental resolution. Contrary to what would be expected for a spin-glass state \cite{Mydosh1993, Bastien2019}, $\chi' (T)$ does not exhibit any frequency dependent maximum. Instead, our data show a strong similarity to the dc magnetic susceptibility with an increasing magnetic susceptibility towards lower temperatures. Thus, our ac susceptibility measurements rule out the possibility of a bulk spin-glass state at least for $T > $~2\,K in Ru$_{0.79}$Rh$_{0.21}$Cl$_3$, while ZFC-FC splitting may indicate a weak glassy behavior, which may arise from a small amount of magnetic impurities in the sample. 

\begin{figure*}[t]
\begin{minipage}{0.49\linewidth}
\begin{center}
\includegraphics[width=1\linewidth]{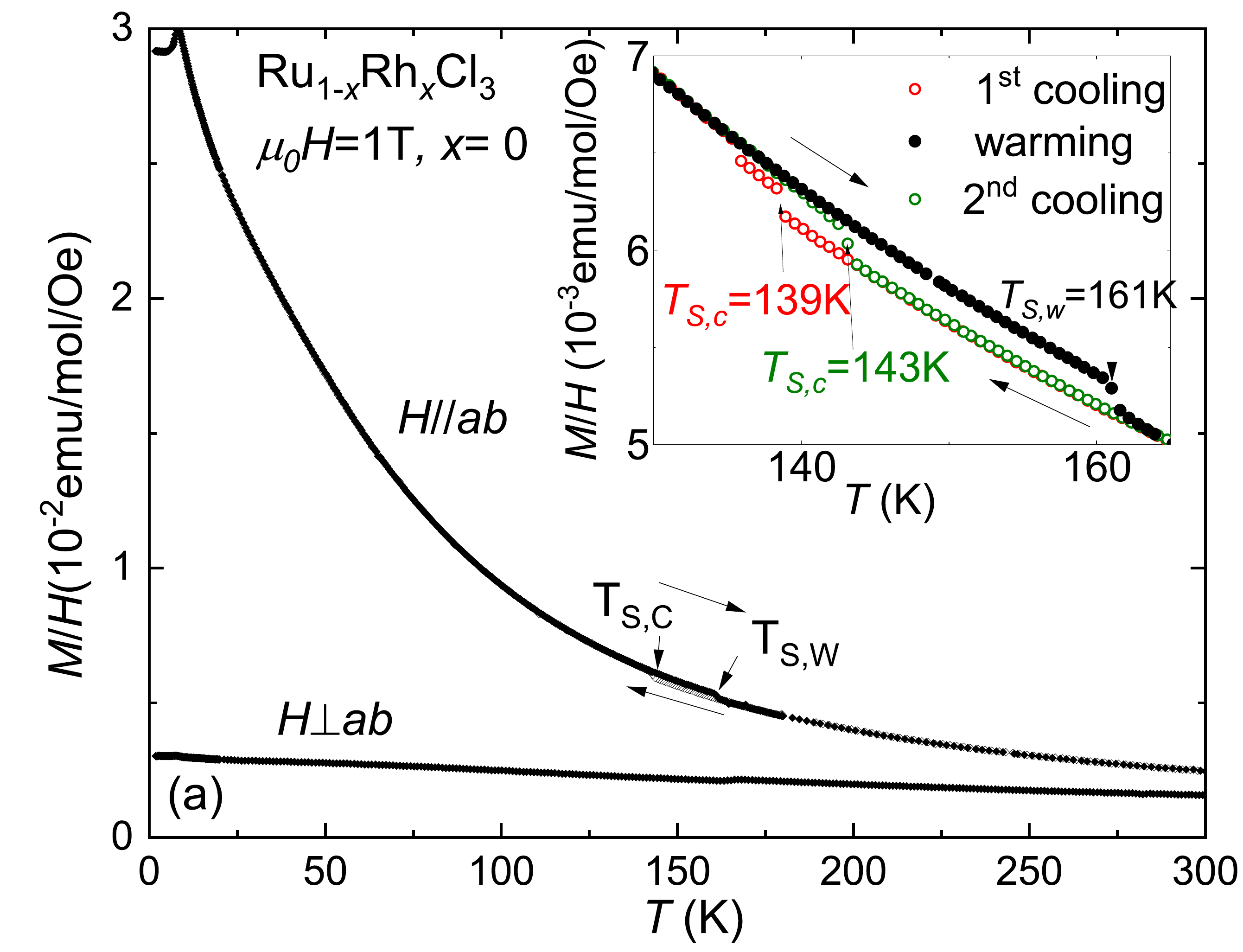}
\includegraphics[width=1\linewidth]{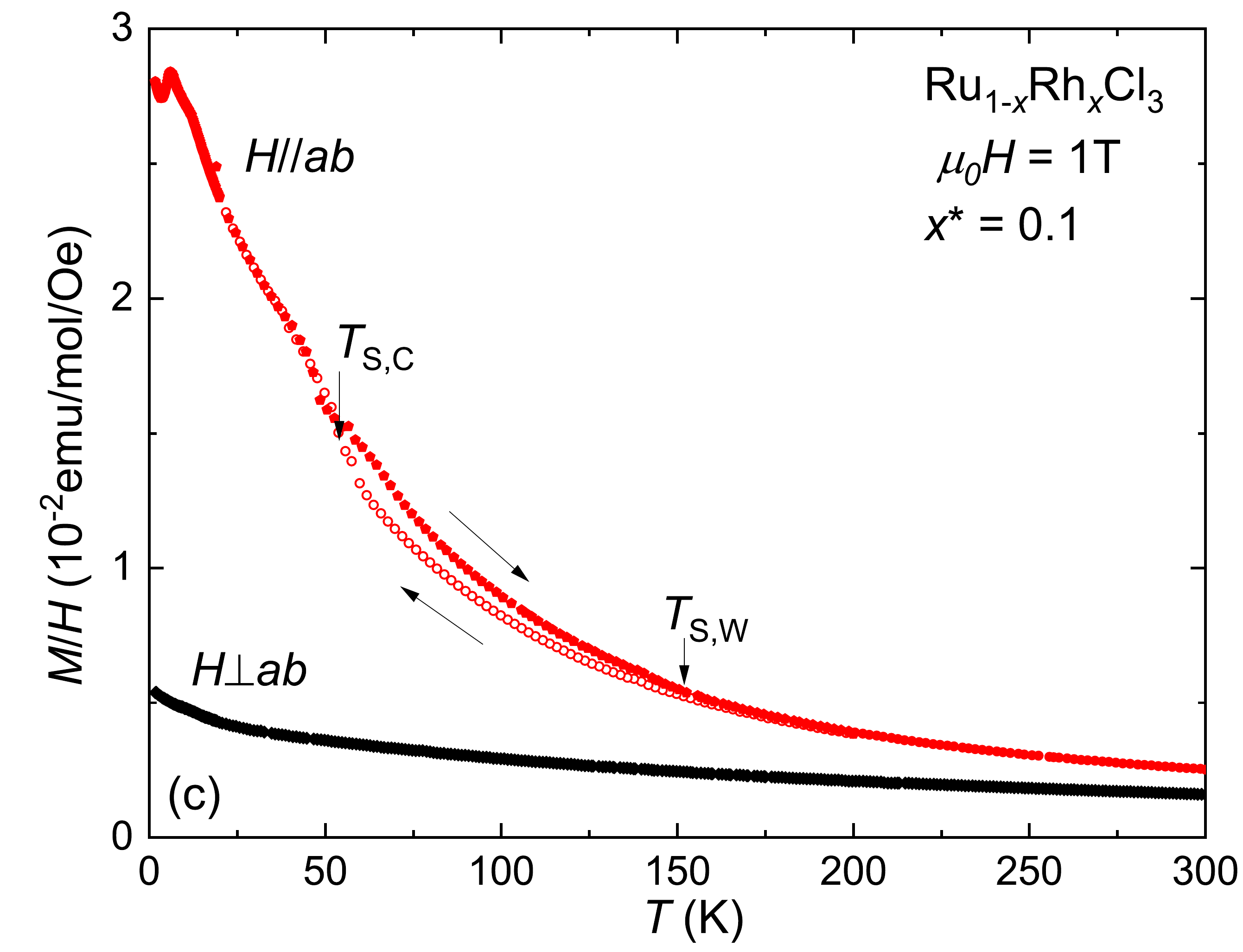}
\end{center}
\end{minipage}
\hfill
\begin{minipage}{0.49\linewidth}
\begin{center}
\includegraphics[width=1\linewidth]{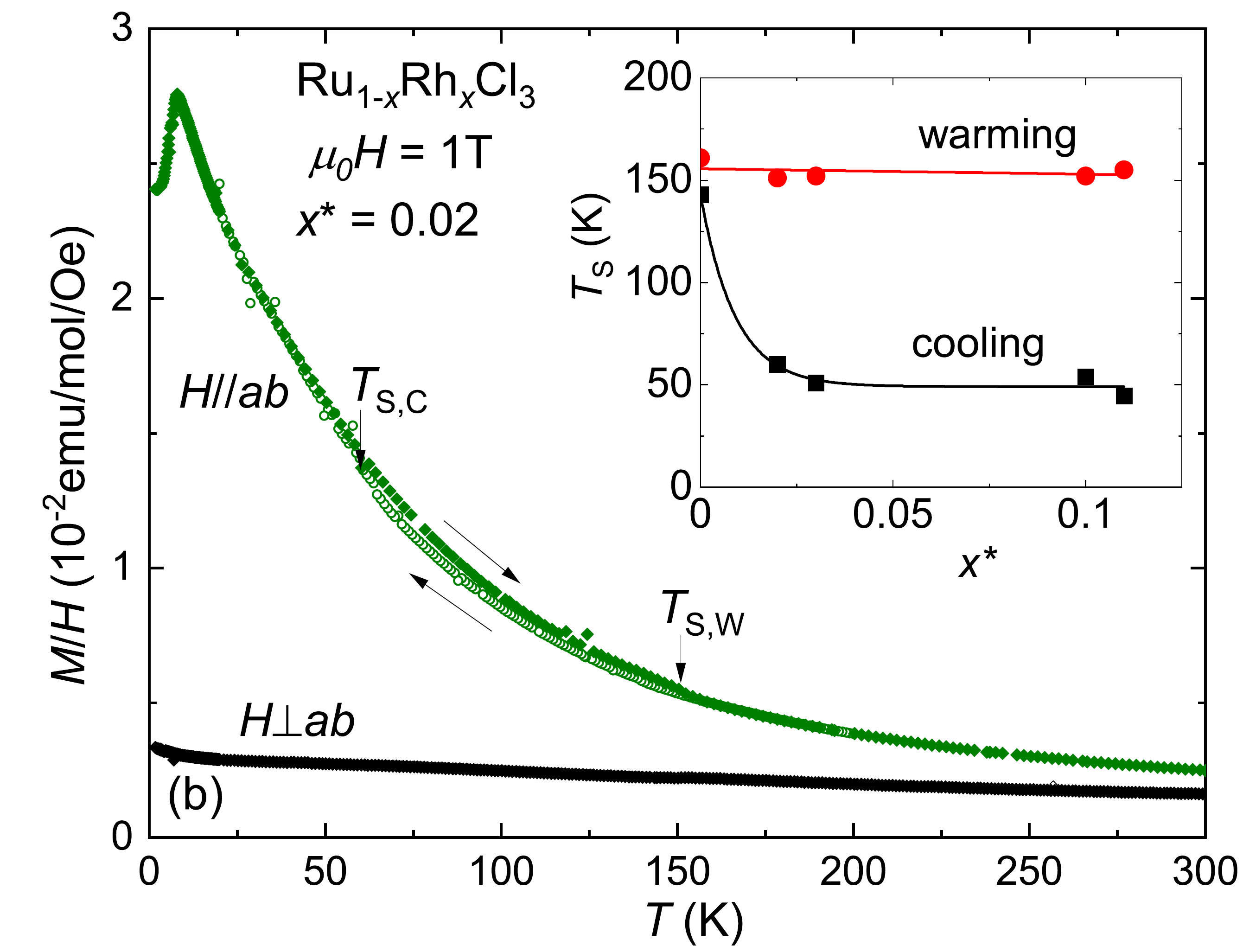}
\includegraphics[width=1\linewidth]{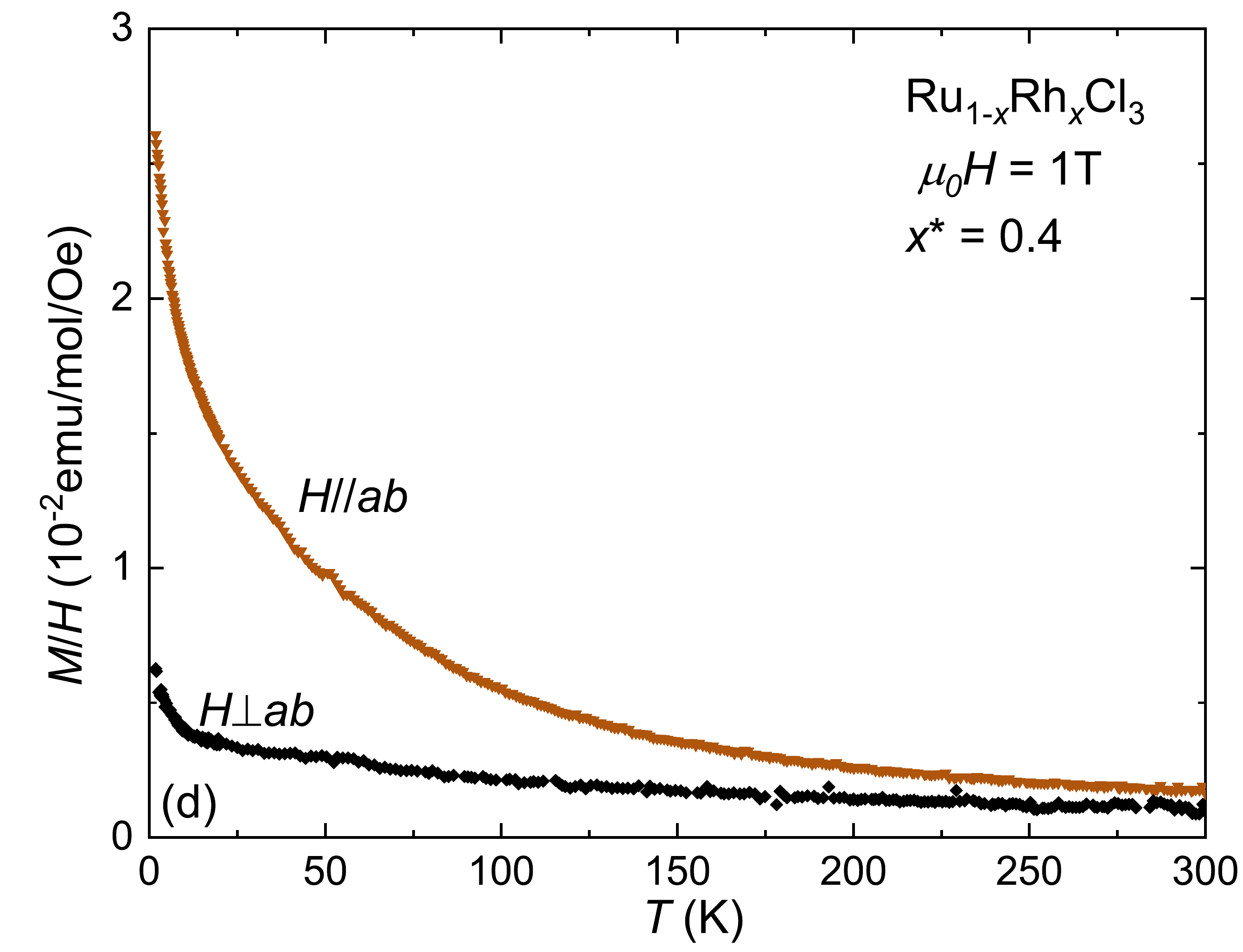}
\end{center}
\end{minipage}
\caption{Magnetization divided by magnetic field $M/H$ of Ru$_{1-x}$Rh$_x$Cl$_3$ as a function of temperature up to 300\,K for (a) $x=0$, (b) $x^*=$0.02, (c) $x^*=$0.1 and (d) $x^*=$0.4. The magnetic field was applied in the $ab$ plane and transverse to the $ab$ plane.  Open and full symbols stand for measurements upon warming and upon cooling, respectively. The hysteresis loop at the structural transition was measured during temperature sweeps with the rate of 1~K/min. The inset in (a) is a zoom on the structural transition showing the difference between measurements upon the first cooling of the sample, upon warming and upon the second cooling. The inset in (b) shows the dependence of the structural transition temperature on the Rh content $x^*$ upon warming and upon the second cooling of the crystal.}
\label{anisotropy}
\end{figure*}

The temperature dependence of the normalized magnetization $M/H$ is represented in Fig.~\ref{anisotropy} for temperature up to 300\,K both for magnetic fields applied in the basal plane $ab$ and transverse to it. The mother compound $\alpha$-RuCl$_3$ shows a strong magnetic anisotropy $\chi_{ab}/\chi_{c}\approx 9$ at $T=15\,$K induced by the anisotropic magnetic interactions $K$ and $\Gamma$ \cite{Majumder2015, Lampen-Kelley2018}. Overall, this anisotropy gets reduced upon substitution together with a reduction of the number of nearest-neighbor bonds to reach  $\chi_{ab}/\chi_{c}\approx 4.5$ at $T$ = 15~K for the substitution ratio $x^* = 0.4$ (Fig.~\ref{anisotropy} (b)-(d)). Previous work on Ru$_{1-x}$Ir$_x$Cl$_3$ reported a non-monotonous evolution of the magnetic anisotropy under substitution with a minimum of the magnetic anisotropy for $x=0.04$~\cite{Do2018}. Such an effect was not resolved in our present study on Ru$_{1-x}$Rh$_x$Cl$_3$.

Our magnetization measurements further clearly indicate a structural transition with a hysteresis in temperature and an increase of the magnetic susceptibility in the $ab$ plane by cooling through it. A slight shift of the structural transition upon cooling towards higher temperatures was observed between the first and the second cooling of the crystal as represented in Fig.~\ref{anisotropy}(a) for $x=0$ ($\alpha$-RuCl$_3$). Similar shifts of the structural transition upon thermal cycling were previously reported in temperature dependent X-ray diffraction studies in the isostructural compound CrCl$_3$~\cite{McGuire2017}. While the hysteresis loop for the second thermal cycle for $\alpha$-RuCl$_3$ is between $T_{S,C}$ = 143\,K and $T_{S,W}$ = 161\,K, it extends for the doped crystal with $x^*=0.02$ down to $T_{S,C} \sim$ 60\,K. The strong sensitivity of $T_{S,C}$ to the chemical composition characterized here sheds light on the discrepancy of previously reported structural transition temperatures upon cooling in pure $\alpha$-RuCl$_3$~\cite{Kubota2015,Park2016,Baek2017,He2018,Widmann2019,Gass2020}. Indeed, several studies reported the occurrence of the structural transition upon cooling in pure $\alpha$-RuCl$_3$ around 60\,K~\cite{Park2016,Baek2017,He2018}, similar to our observation in Ru$_{1-x}$Rh$_x$Cl$_3$ for $0.02 \le x^* \le 0.11$. Note that the shift of the structural transition temperature towards lower temperature under chemical substitution in Ru$_{1-x}$Rh$_x$Cl$_3$ may either be a consequence of the change of the lattice parameters upon substitution or a consequence of local disorder. 

\begin{figure}
\begin{minipage}{0.5\linewidth}
	\begin{center}
		\includegraphics[width=1\linewidth]{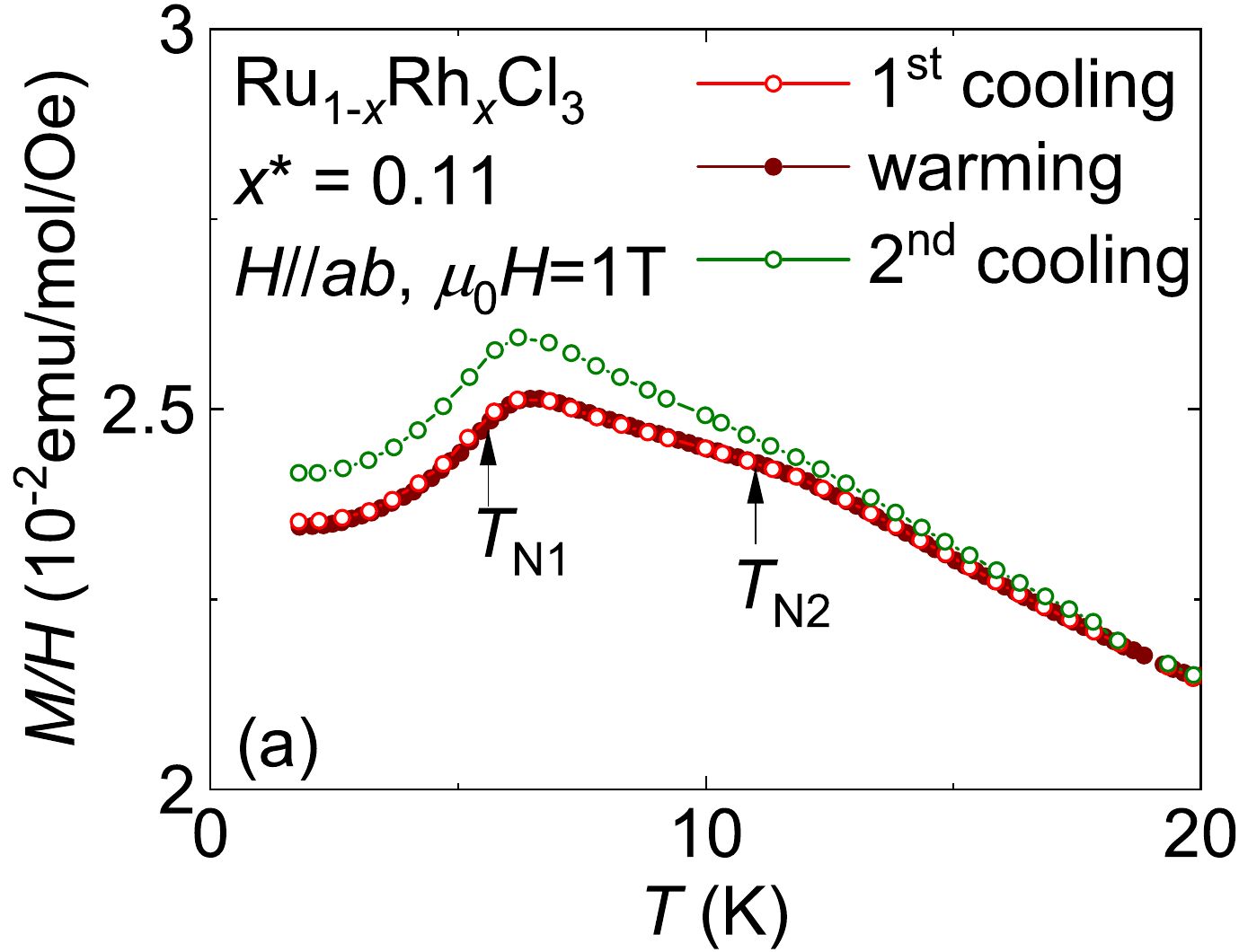}
	\end{center}
\end{minipage}\begin{minipage}{0.5\linewidth}
\begin{center}
\includegraphics[width=1\linewidth]{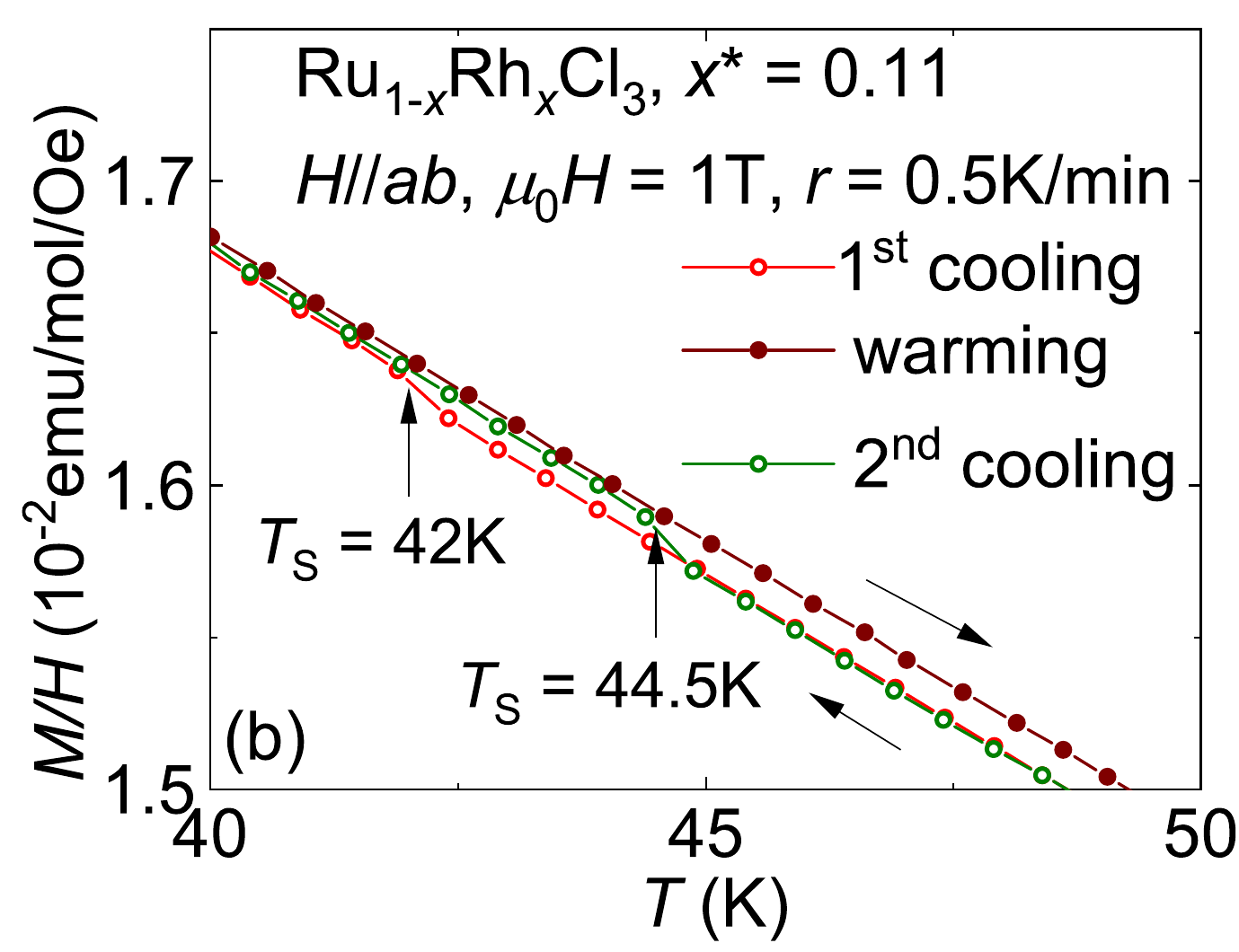}
\end{center}
\end{minipage}
\begin{minipage}{0.5\linewidth}
	\begin{center}
		\includegraphics[width=1\linewidth]{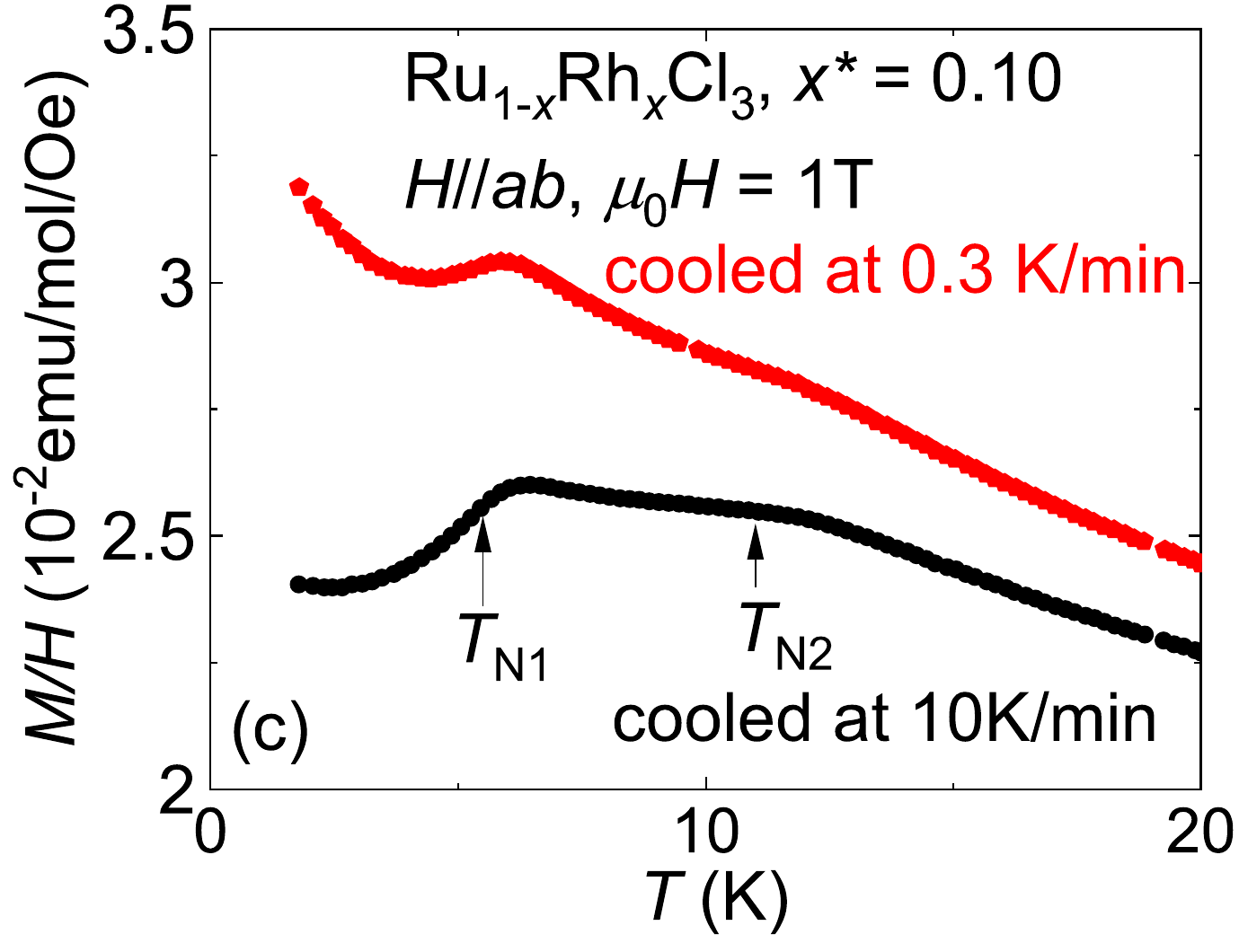}
	\end{center}
\end{minipage}\begin{minipage}{0.5\linewidth}
\begin{center}
\includegraphics[width=1\linewidth]{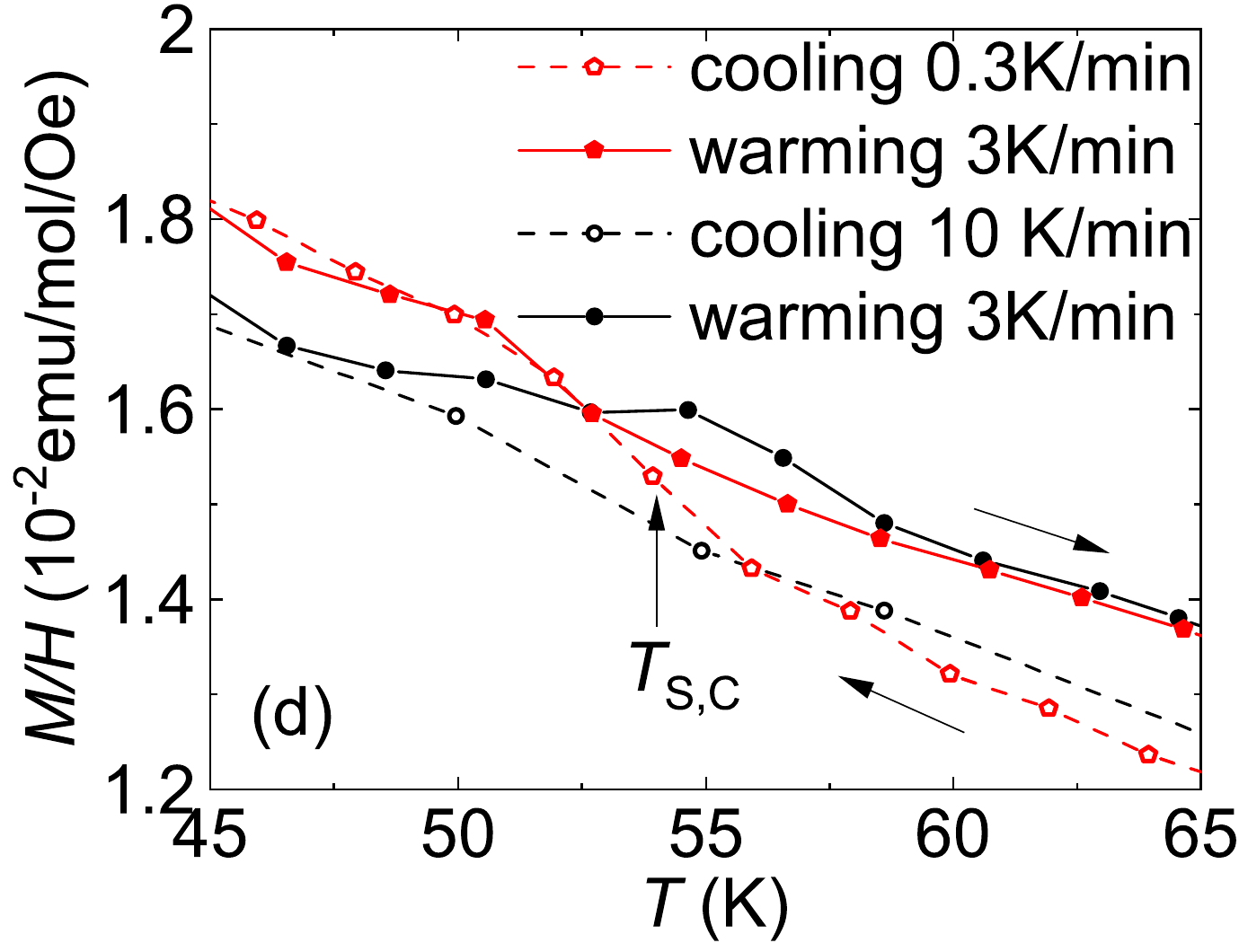}
\end{center}
\end{minipage}\begin{center}
\caption{(a) and (b) Normalized magnetization of the  $x^*=0.11$ crystal in the $ab$ plane measured after a single cooling through the structural transition at 0.5\,K/min and remeasured after having warmed the crystal back to 200\,K at 0.5\,K/min and cooled it a second time through the structural transition at 0.5\,K/min. (c) and (d) Normalized magnetization of Ru$_{1-x}$Rh$_x$Cl$_3$ for $x^*=0.1$ measured after a cooling through the structural transition at 0.3\,K/min and remeasured after having warmed the crystal back to 200\,K and cooled it through the structural transition at about 10\,K/min. Measurements upon warming were performed with the rate of 0.3\,K/min below 20K and 3\,K/min above 20K.}
\label{TS}
\end{center}
\end{figure}

The magnetization of Ru$_{1-x}$Rh$_x$Cl$_3$, $x^*=0.11$ was measured as a function of temperature upon successive thermal cycling with a low sweep rate through the structural transition of 0.5~K/min and it is represented in  Fig.~\ref{TS}(a-b). A magnetic field of $\mu_0H$=1~T was applied for precise magnetization measurements.
The effect of this magnetic field on the structural transition can be neglected since the contraction of the lattice parameter $c$ induced by a magnetic field of $\mu_0H$=1~T in the paramagnetic state is more than three orders of magnitude lower than the contraction occurring at the structural transition~\cite{Gass2020, Kocsis2022}.
In addition to the shift of the structural transition towards higher temperatures, we observe a slight increase of the magnetization in the paramagnetic state and a reduction of the kink at $T_{\mathrm N2}$. This indicates a reduction of the amount of stacking faults upon successive thermal cycling. We performed a different procedure with a crystal of similar composition Ru$_{1-x}$Rh$_x$Cl$_3$, $x^*=0.1$ and present the magnetization as a function of temperature in Fig.~\ref{TS}(c-d). The crystal was cooled through the structural transition at 0.3\,K/min, then warmed up and cooled through the structural transition with a rate of 10\,K/min. The effect is opposite compared to the previous case with a slow cooling rate: the low-temperature magnetization after the second cooling is lower than after the first cooling and the kink at $T_{\mathrm N2}$ is also magnified after the second cooling. This indicates that a fast cooling through the structural transition multiplies the stacking faults and in turn, that the structural phase transition is incomplete.

This shift of the structural transition temperature towards lower temperature under chemical substitution leads to an incomplete structural transition upon cooling due to slow dynamics of the structure around 50\,K. It implies a slight variation of the crystal structure depending on the history of thermal cycling of the crystal. The observation of a (rigorous) dependence of the magnetization on the history of thermal cycling confirms that the low-temperature magnetic properties of $\alpha$-RuCl$_3$ and its doping variants strongly depend on the details of the crystal structure and in particular on the stacking of the honeycomb layers. It is important to notice that the evolution of the magnetization upon thermal cycling does not imply any shift of the magnetic transitions but only a change of the relative magnitude of the two different magnetic transitions. Thus this effect does not prevent us from drawing the intrinsic $(x^*-T)$ phase diagram of Ru$_{1-x}$Rh$_x$Cl$_3$ (see Fig. \ref{MvsT}(d)). 

\subsection{Specific heat measurements}

\begin{figure}[t]
\begin{center}
\includegraphics[width=0.9\linewidth]{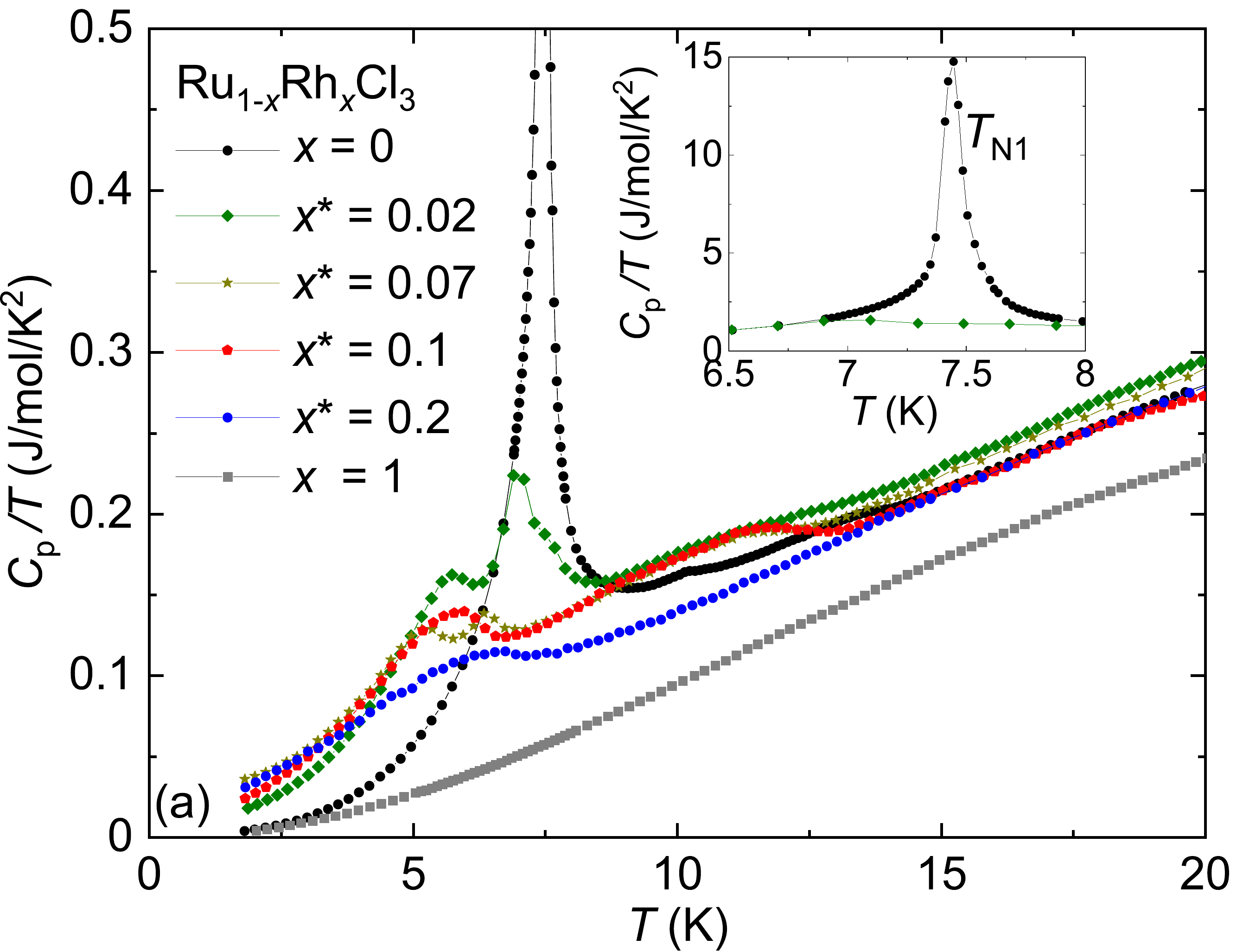}
\includegraphics[width=0.9\linewidth]{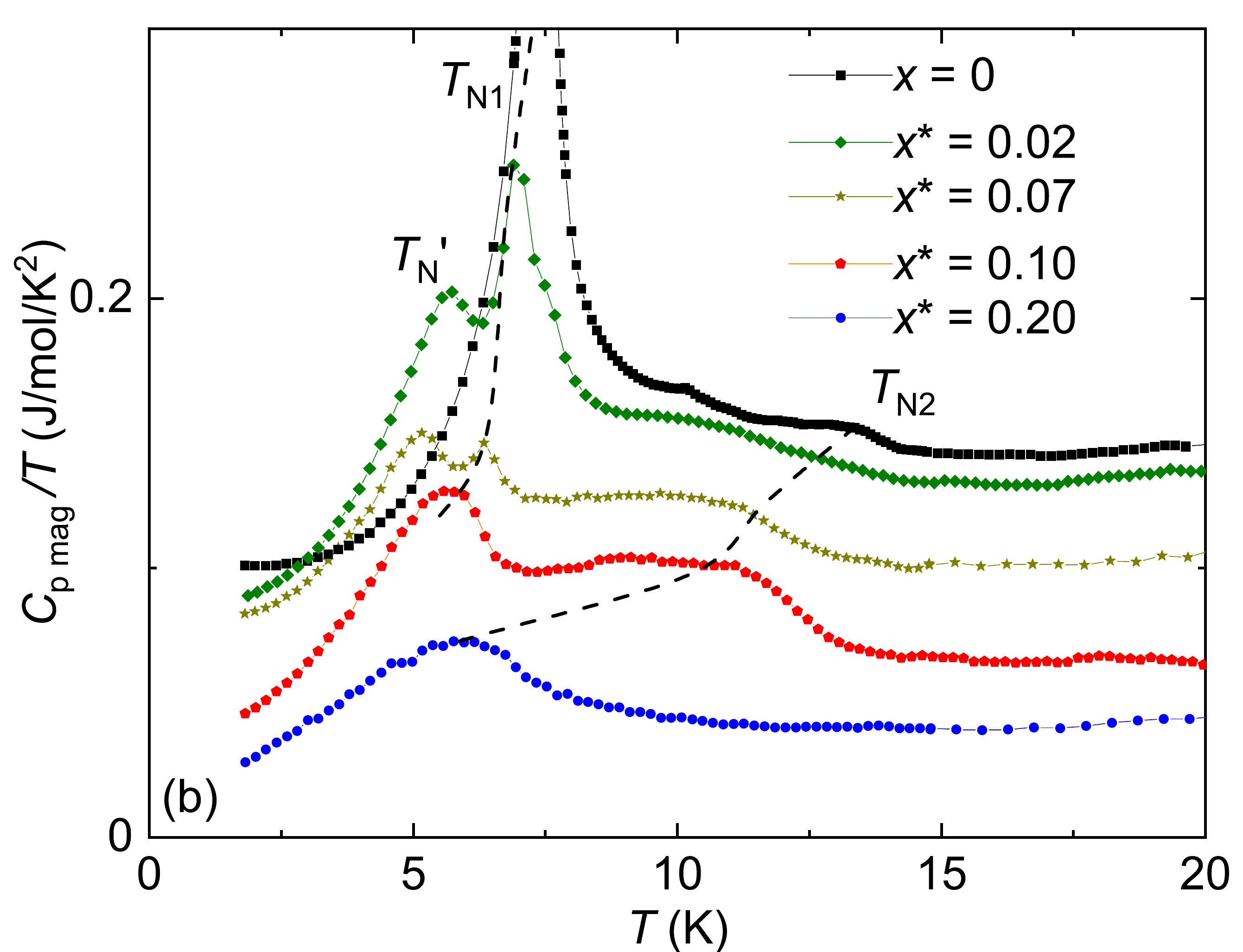}
\caption{(a) Specific heat divided by temperature of the Ru$_{1-x}$Rh$_x$Cl$_3$ series as a function of temperature. The data for $x=0$ and  $x^*=0.02$ are represented in the inset on a different scale to show  the sharp and high peak at the magnetic transition for $x=0$. (b) Magnetic contribution to the specific heat divided by the temperature $C_{p,\rm{mag}}/T$ as a function of temperature. The data for $x^* >$ 0 are offset for clarity and the arrows indicates the shift of the magnetic transitions $T_{\mathrm N1}$ and $T_{\mathrm N2}$.}
\label{Cp}
\end{center}
\end{figure}

The specific heat coefficients $C_p/T$ of some representative samples of the Ru$_{1-x}$Rh$_x$Cl$_3$ series are represented in Fig.~\ref{Cp} as a function of temperature up to $T$ = 20~K. The data for $x=0$ and $x=1$ were previously published in Ref.~\onlinecite{Wolf2022} and~\onlinecite{Wolter2017} respectively. The effect of successive thermal cycling on the specific heat was not investigated and the crystals were cooled  at 1\,K/min through the structural transition.
Note that the endmember RhCl$_3$ serves as a non-magnetic analog to estimate the phononic contribution to the specific heat. Its specific heat was subtracted from the specific heat of Ru$_{1-x}$Rh$_x$Cl$_3$  to extract the magnetic contribution to the specific heat without the need of any scaling~\cite{Bastien2019}. 

The resulting magnetic contribution $C_{p,\rm{mag}}/T$ of Ru$_{1-x}$Rh$_x$Cl$_3$ is represented in Fig.~\ref{Cp}(b) as a function of temperature. Remarkably, $C_{p,\rm{mag}}/T$ is still sizeable at $T$ = 20\,K for all samples shown in Fig.~\ref{Cp} (b) in agreement with large spin fluctuations reaching up to temperatures of the exchange couplings of the order of 50\,K and more. 
In addition to the main peak at $T_{\mathrm N1}$ signaling the antiferromagnetic transition, a broad bump around $T_{\mathrm N2}$ appears, which for some of the samples was not previously detected by the magnetization measurements. This second transition is known as a signature of a progressive ordering associated to stacking faults prior to the long-range magnetic ordering at $T_{\mathrm N1}$ ~\cite{Cao2016}. By defining the magnetic ordering temperature at the maximum of $C_p/T$, we are able to follow the evolution of $T_{\mathrm N1}$ and $T_{\mathrm N2}$ with Rh content (Fig.~\ref{Cp}(b)). The temperatures thus obtained are in good agreement with the ones observed in magnetization measurements (Fig.~\ref{MvsT}(d)).

Furthermore, the specific heat measurements uncover a third transition $T_{\mathrm N}'$ occurring at slightly lower temperatures than $T_{\mathrm N1}$ for $x^*=0.02$ and 0.07. This additional transition may have various origins. The temperature range in which it appears suggests that it is probably not related to stacking faults, which usually induce a magnetic transition at temperatures around $T_{\mathrm N2}\approx$ 10--14\,K. 

One possibility is that it originates from a region of the crystal with higher substitution rate, which cannot be distinguished from our thorough structural characterization. On the other hand, previous reports indicate that two successive magnetic transitions are typical for some antiferromagnets with triangular, honeycomb or Kagom\'{e} lattices \cite{Matsubara,Lee_2016,Maegawa,Sannigrahi}. Driven by frustration and anisotropy, incommensurate to commensurate transitions or spin re-orientations of the magnetic lattice have been observed. While Kitaev magnets exhibit unique properties beyond geometrical frustration, they also show a large variety of competing ground states, leaving the possibility of similar successive magnetic phase transitions. Possible changes on the magnetic structure for temperatures between $T_{\mathrm N1}$ and $T_{\mathrm N}'$ might be addressed in future neutron scattering studies.

Then, these two transitions merge in a broad one around 5\,K for $x^*=0.1$. The measurement for $x^*=0.2$ was performed on a crystal harboring the magnetic ordering at $T_{\mathrm N2}$. We could not perform reliable measurements of the specific heat of crystals with $x^*>$ 0.2, which do not show any magnetic order in magnetization measurements, because of their rather low mass of $m<0.3$\,mg.

\begin{figure*}[t]
\begin{minipage}{0.49\linewidth}
\begin{center}
\includegraphics[width=1\linewidth]{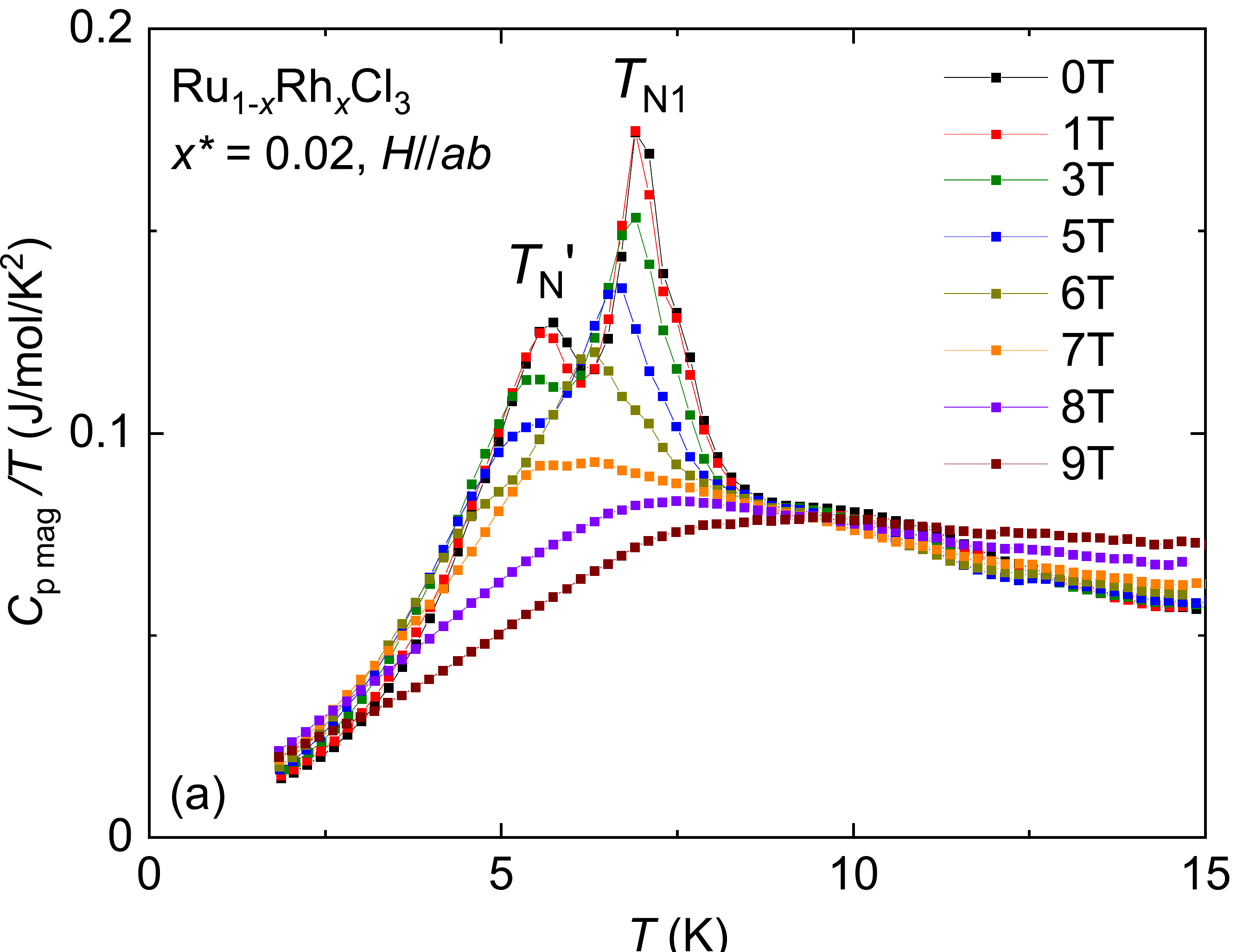}
\includegraphics[width=1\linewidth]{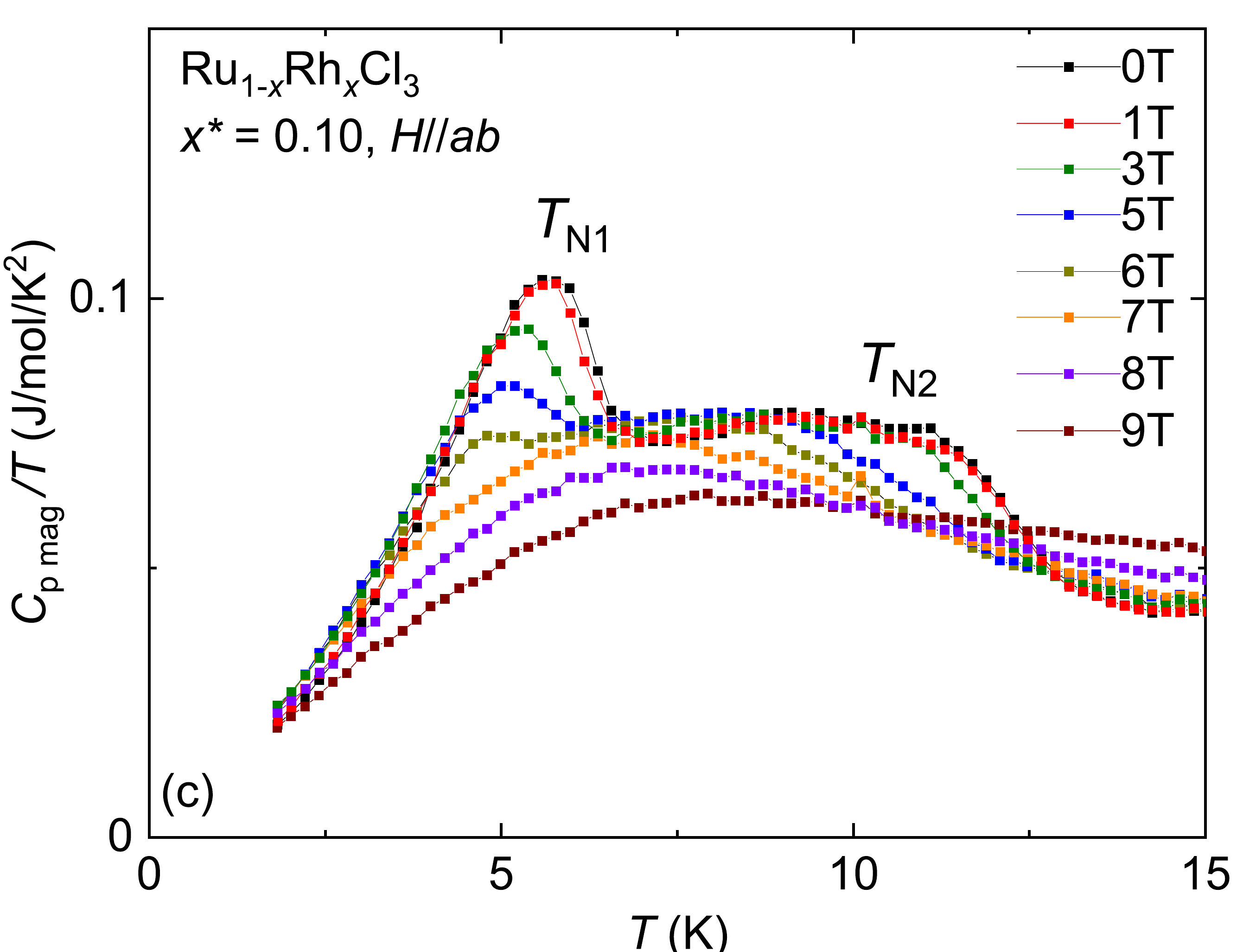}
\end{center}
\end{minipage}
\hfill
\begin{minipage}{0.49\linewidth}
\begin{center}
\includegraphics[width=1\linewidth]{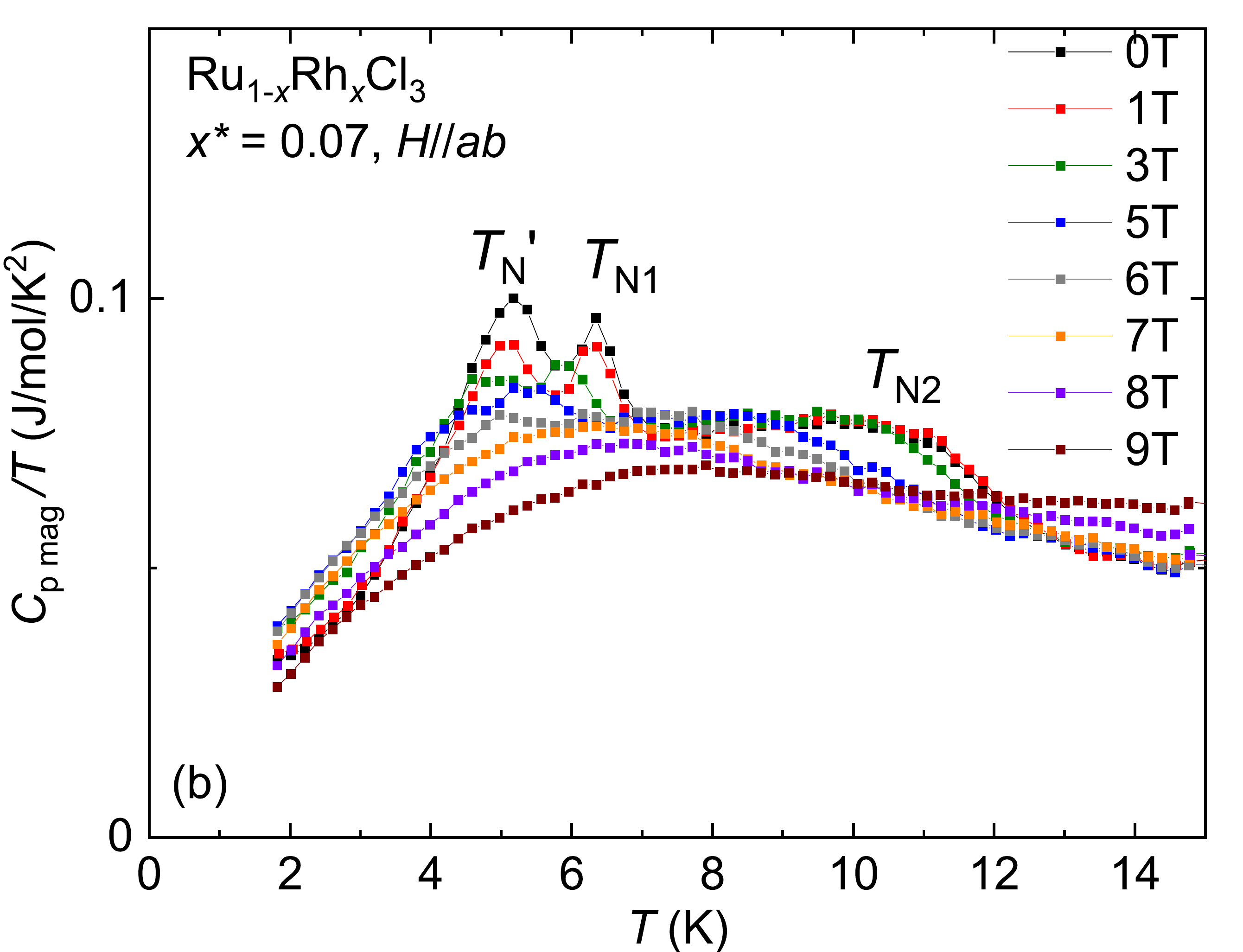}
\includegraphics[width=1\linewidth]{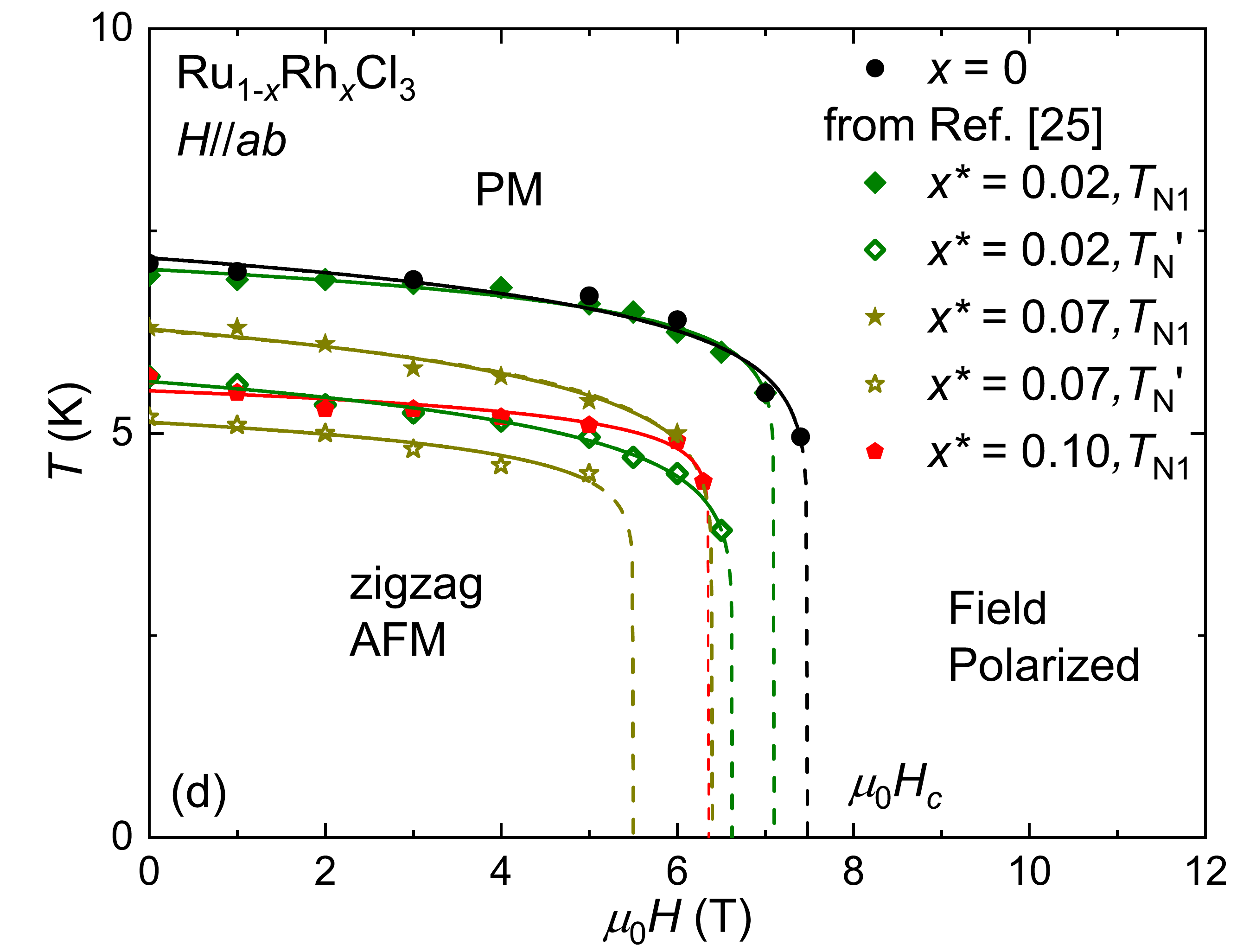}
\end{center}
\end{minipage}
\caption{(a-c) Magnetic contribution to the specific heat divided by temperature $C_{p,\rm{mag}}/T$ for  $x^*=0.02$,  $x^*=0.07$, and  $x^*=0.1$, respectively, for various magnetic fields applied in the basal plane $ab$. (d) Magnetic-field temperature phase diagram of Ru$_{1-x}$Rh$_x$Cl$_3$ for magnetic fields applied in the basal plane $ab$. The data for $x=0$ are taken from Ref.~\onlinecite{Gass2020}. Lines are guides to the eye. For $x^*=0.02$ and $x^*=0.07$ , the full and open symbols correspond to the two successive transitions $T_{\mathrm N1}$ and $T_{\mathrm N}'$.}
\label{HT}
\end{figure*}

The magnetic contribution of the specific heat divided by the temperature $C_{p,\rm{mag}}/T$ of Ru$_{1-x}$Rh$_x$Cl$_3$ for magnetic fields applied in the basal plane $ab$ is represented in Fig.~\ref{HT} together with the subsequent field-temperature phase diagram of Ru$_{1-x}$Rh$_x$Cl$_3$. For $x^*=0.02$, the two successive magnetic transitions $T_{\mathrm N}'$ and $T_{\mathrm N1}$ are suppressed by magnetic fields of $\mu_0H_c'\approx 6.5$\,T and $\mu_0H_c\approx 7$\,T, respectively, above which only a broad maximum of $C_{p,\rm{mag}}/T (T)$ remains. For even higher values of $x^*$ the critical field shifts to even lower values. The possible occurrence of a disordered field-induced spin liquid state in Ru$_{1-x}$Rh$_x$Cl$_3$ in a finite field interval above the critical field $H_c$ remains under debate and needs further investigations on large single crystals of this series via additional microscopic methods~\cite{Baek2017, Kasahara2018, Balz2019, Zhao2022}. Note that there are two possible disordered magnetic phases in Ru$_{1-x}$Rh$_x$Cl$_3$ - the one for substitution ratios higher than $x^*=0.2$ and the one for magnetic fields higher than $H_c$. They may be connected or of different nature.

\section{Discussion}

Our experimental study of the magnetic properties of Ru$_{1-x}$Rh$_x$Cl$_3$ shows a clear suppression of magnetic order of $\alpha$-RuCl$_3$ upon the dilution of the magnetic lattice. The critical concentration to suppress magnetic order in $ABC$-stacked crystals, $x_c\approx 0.2$, is relatively high, indicating that the magnetic ground state of $\alpha$-RuCl$_3$ is relatively robust. In particular, it is higher than for the other substitution series realizing a dilution of the magnetic lattice Ru$_{1-x}$Ir$_x$Cl$_3$ with $x_{c1}=0.13$~\cite{Lampen-Kelley2017, Do2018}. The difference between these two substitution series may arise from slight changes of the crystal structure upon substitution. 

In the case of the substitution of Ru by Rh, the dilution of the magnetic lattice is accompanied by a contraction of the unit cell along each crystallographic axis comparable to the application of hydrostatic pressure. This contraction may contribute to the suppression of the zigzag magnetic order, since hydrostatic pressure was found to reduce the magnetic ordering temperature in the low-pressure limit~\cite{Wolf2022}. However, compressibility measurements in $\alpha$-RuCl$_3$ are missing to compare the effect of chemical substitution on the crystal structure with the results of applied pressure. We can nevertheless notice that the pressure-induced structural transition towards a valence bond crystal at $p_s=0.1$~GPa~\cite{Cui2017,Bastien2018,Biesner2018,Wolf2022} in $\alpha$-RuCl$_3$ is not observed at ambient pressure in Ru$_{1-x}$Rh$_x$Cl$_3$ despite the contraction of the honeycomb layer.

On the contrary, in the case of the substitution of Ru$^{3+}$ by Ir$^{3+}$ the honeycomb layers expand and the interplanar distance $c$ is reduced~\cite{Lampen-Kelley2017}. This deformation is similar to the one induced by uniaxial pressure along the $c$ axis. Such uniaxial pressure was predicted to favor the quantum spin liquid state with respect to the zigzag order~\cite{Kim2016,Kaib2021,Kocsis2022}. Thus, the rapid suppression of the zigzag magnetic order in Ru$_{1-x}$Ir$_x$Cl$_3$ is probably a combined effect of the magnetic dilution and of the specific change of the cell parameters upon substitution.

An open question is whether the disordered quantum magnetic phase occurring in a strongly disordered magnetic lattice such as Ru$_{1-x}$Rh$_x$Cl$_3$ for $x^*>0.2$ can be described as quantum spin liquid phase. The theoretical studies of the diluted Kitaev model in the large dilution limit ($x\approx 0.2$) predicted a Majorana spin liquid with emergent magnetic moments implying a large paramagnetic tail~\cite{Sanyal2021,Nasu2021}. This phase might be present in Ru$_{1-x}$Rh$_x$Cl$_3$ for $x^*>0.2$ despite the additional interactions $J$ and $\Gamma$, however, it would be difficult to identify whether the experimental paramagnetic tail comes from emergent quantum magnetic moments or from localized classical moments. Note that the difficulty to identify quantum spin liquid phases in disordered systems goes beyond the Kitaev model and it was also recently discussed in triangular magnets such as YbMgGaO$_4$~\cite{Li2017, Kimchi2018} and NaYb$_{1-x}$Lu$_x$S$_2$~\cite{Haeussler2022}, in kagome magnets such as Tm$_3M_2$Sb$_3$O$_{14}$ (M = Mg, Zn)~\cite{Ma2020} and in double perovskites such as Sr$_2$CuW$_{1-x}$Te$_{x}$O$_6$~\cite{Mustonen2018}.

Previous works have shown that the structural transition of $\alpha$-RuCl$_3$ from the monoclinic to the rhombohedral lattice upon cooling is incomplete, leading to stacking faults responsible for multiple magnetic transitions~\cite{Cao2016,He2018,Lampen-Kelley2018}. In this work, we pointed out a shift of the structural transition upon cooling $T_{S,C}$ towards higher temperature with successive thermal cycling, which shows that the room-temperature monoclinic structure also keeps the memory of the low-temperature structure after warming of the crystal. It seems that the structural transition induced upon cooling is not completely reversed by warming up the crystal. In the case of the Kitaev magnet $\alpha$-RuCl$_3$, the large magnetoelastic coupling gives the possibility to probe such effects via magnetic measurements. This effect becomes visible in the substituted crystals Ru$_{1-x}$Rh$_x$Cl$_3$ thanks to the shift of the structural transition towards lower temperature.

\section{Summary and conclusion}
We report the single-crystal growth of the Ru$_{1-x}$Rh$_x$Cl$_3$ series ($x$ = 0 - 0.6) and the characterization of its structural and magnetic properties. The chemical substitution leads to the dilution of the magnetic lattice. From a structural point of view, the honeycomb lattice of transition metal cations is preserved in the series with statistical distribution of Ru and Rh, and the monoclinic lattice shrinks linearly with increasing Rh content. The zigzag magnetic ground state of $\alpha$-RuCl$_3$ remains up to a substitution ratio around $x^*\approx 0.2$, and the magnetic field-induced transition towards a possible quantum spin liquid state shifts to slightly lower magnetic fields for $x^*\approx 0.1$. These results show that the zigzag magnetic order of $\alpha$-RuCl$_3$ is relatively robust upon magnetic dilution. We propose that chemical pressure would also contribute to the suppression of the zigzag magnetic order, especially in the previously reported substitution series Ru$_{1-x}$Ir$_x$Cl$_3$.

The magnetically disordered phase obtained for $x^*> 0.2$ does not show any signature of freezing into a spin glass. An open question is whether such a quantum disordered magnetic phase occurring on a strongly disordered magnetic lattice can (conceptually) be described as a quantum spin liquid phase, leaving room for interesting future discussions and work within the community. We also demonstrated that cooling rates as well as successive thermal cycling must be considered for precise studies of structural and magnetic properties of van-der-Waals magnets due to the possible occurrence of irreversible modifications of the structural properties in case of crossing structural phase transitions.


\begin{acknowledgments}

G.B. and E.V. contributed equally to this work.
Technical support in magnetization measurements by S. Ga\ss \, (IFW Dresden) is gratefully acknowledged. We are indebted to Dr.~Ph.~Schlender and Prof.~J.~Weigand (Faculty of Chemistry and Food Chemistry, TU Dresden, Germany) for enabling the low-temperature SCXRD data acquisition and for assistance in the data processing and refinement. We are very grateful to Mr.~S.~Seker and Ms.~M.~Trokoz for their contributions to synthesis, and to MSc.~A.~Brunner for the know-how preparation of chlorine capillaries. We thank L. Janssen, M. Vojta (TU Dresden), D. Hovan\v{c}\'ik, M. Kratochv\'ilov\'a, J. Posp\'i\v{s}il, and V. Sechovsk\'y (Charles University) for insightful discussions.
We acknowledge financial support from the DFG through SFB 1143 (project-id 247310070) and the W\"urzburg-Dresden Cluster of Excellence on Complexity and Topology in Quantum Matter -- \textit{ct.qmat} (EXC 2147, project-id 390858490) as well as from the European Union's Horizon 2020 research and innovation programme under the Marie Sk\l{}odowska-Curie grant agreement No 796048.  G. B. acknowledges financial support from the Primus Research Programme  of the Charles University in Prague, Czech Republic. L.T.C. is funded by the DFG (project-id 456950766).

\end{acknowledgments}

\bibliographystyle{apsrev4-1}
\bibliography{RuRhCl}

\end{document}